\documentclass[11pt]{article}
\usepackage[left=1in,top=1in,right=1in,bottom=1in,head=0in,nofoot]{geometry}

\usepackage{setspace,url,bm,amsmath} 
\usepackage{scalerel}

\usepackage{xcolor}
 \usepackage{multirow}
 \usepackage{arydshln}
 \usepackage{mathtools}
\usepackage{titlesec} 
\titlelabel{\thetitle.\quad} 
\titleformat*{\section}{\bf\Large\center}

\usepackage{authblk}
\usepackage{float}
\usepackage{graphicx} 
\usepackage{bbm}
\usepackage{latexsym}
\usepackage{caption}
\usepackage[margin=20pt]{subcaption}
\usepackage{enumerate}

\usepackage{amsthm}
\usepackage{amssymb}
\usepackage{amsmath}
\usepackage{color}

\usepackage{comment}
\theoremstyle{definition}
\newtheorem{assumption}{Assumption}
\newtheorem*{theorem*}{Theorem}
\newtheorem{theorem}{Theorem}
\newtheorem*{rmk*}{remark}
\newtheorem{proposition}{Proposition}
\newtheorem{lemma}{Lemma}

\newtheorem{condition}{Condition}

\newtheorem{remark}{Remark}

\newtheorem*{corollary*}{Corollary}

\usepackage{natbib} 
\bibpunct{(}{)}{;}{a}{}{,} 

\usepackage{etoolbox} 
\apptocmd{\sloppy}{\hbadness 10000\relax}{}{} 

\usepackage{multibib}
\newcites{sec}{References}

\usepackage{color}
\usepackage{listings}
\usepackage{hyperref}
\usepackage{booktabs}

\usepackage{lscape}

\RequirePackage[normalem]{ulem}

\def \P{\mathbb{P}} 
\def \V {\mathbb{V}}

\def \E {\mathbb{E}}

\def \V {\mathbb{V}}
\def \I {\mathbb{I}}
\def \cI {\mathcal{I}}

\newcommand{\indep}{\perp \!\!\! \perp}

\usepackage{algpseudocode}
 \usepackage{algorithm}

\allowdisplaybreaks

\begin{document}

\singlespacing

\title{\bf Safe Individualized Treatment Rules  \\ with Controllable Harm Rates}

\author[1]{Peng Wu}
\author[2]{Qing Jiang}
\author[1]{Shanshan Luo\thanks{Corresponding author: shanshanluo@btbu.edu.cn}}
\author[1,3]{Zhi Geng}
\affil[1]{\small School of Mathematics and Statistics, Beijing Technology and Business University}
\affil[2]{\small Faculty of Arts and Sciences, Beijing Normal University}
\affil[3]{\small School of Mathematical Sciences, Peking University}



\date{}

\maketitle

\begin{abstract}
Estimating individualized treatment rules (ITRs) is crucial for tailoring interventions in precision medicine. Typical ITR estimation methods rely on conditional average treatment effects (CATEs) to guide treatment assignments. However, such methods overlook individual-level harm within covariate-specific subpopulations, potentially leading many individuals to experience worse outcomes under CATE-based ITRs. In this article, we aim to estimate ITRs that maximize the reward while ensuring that the harm rate induced by the ITR remains below a pre-specified threshold. We first derive the explicit form of the oracle ITR. However, the oracle ITR is not achievable without strong assumptions, as the harm rate is generally unidentifiable due to its dependence on the joint distribution of potential outcomes. 
To address this, we propose two strategies for estimating ITRs with a harm rate constraint under partial identification and establish their large-sample properties.  
By accounting for both reward and harm, our method provides a reliable solution for developing ITRs in high-stakes domains where harm is a critical consideration.  Extensive simulations demonstrate the effectiveness of the proposed methods in controlling harm rates. 
We apply the proposed method to analyze two real-world datasets from a new perspective,  assessing the potential reduction in harm rate compared with historical interventions.  
\end{abstract}



\newpage

\onehalfspacing


\section{Introduction}  

Personalized medicine has gained significant attention in modern biomedical research, aiming to tailor medical treatments to individuals based on their characteristics~\citep{Kosorok+Laber:2019}. 
To this end, numerous statistical methods have been developed to estimate individualized treatment rules (ITRs) using data from clinical trials or observational studies~\citep[see e.g.,][]{Qian2011AoS, Zhao-etal2012,  zhang2012robust, Chen-etal2016, Zhou02012017, zhao2019efficient, Qiu-etal2021, Pu-Zhang2021, Kallus2021,  Zhang-etal2024}. 
Beyond medicine, the estimation of ITR has broad applications across various fields, such as recommending the most effective training programs for the unemployed \citep{Wunsch2013}, identifying optimal strategies to boost voter turnout \citep{Imai-Ratkovic2013}, and assisting policymakers in selecting welfare program eligibility policies \citep{Sun2024}. 
Harmlessness is a critical consideration of ITR estimation, particularly in high-stakes settings.  In clinical medicine, the Hippocratic Oath guides practitioners to uphold the principle of ``do no harm"~\citep{Jonsen1978, morath2005no, Wiens-etal2019}. Similarly, in public policy, decision-makers must carefully evaluate the potential harms of their policies to prevent unintended negative consequences for the population~\citep{Parsons1996PublicPA}. Following \citet{2022nathan}, 
\emph{an ITR is considered harmless (or harmful) for an individual if the assigned treatment yields an outcome that is no worse (or worse) than  it would have been without treatment.}
The harm rate induced by an ITR is the proportion of individuals harmed under its assignment, 
Despite its importance, harmlessness in ITR estimation remains underexplored. The typical framework for estimating ITRs focuses on maximizing the reward (i.e., the average outcome, assuming a larger outcome is better) if all  individuals were to receive the treatment assigned by the ITR. In this framework, the oracle ITR essentially relies on the sign of conditional average treatment effects (CATEs) to determine whether to assign treatment~\citep{2018Who, Athey-Wager2021}.     
However, this strategy maximizes the reward only within subpopulations defined by covariates, while overlooking the potential harm to individuals~\citep{2022nathan}, which may result in overly aggressive ITRs that compromise individual safety. This limitation arises because CATE captures heterogeneity at the subgroup level but fails to reflect individual-level risk~\citep{Wu-etal-2024-Harm}. 
For example, in one subpopulation, a drug benefits 80\% of individuals but harms 20\%, while in another, it benefits 60\% and has no effect on the rest. Despite both subpopulations having a positive CATE of 0.6, assigning the drug to the first subpopulation would still result in harm. 

A notable real-world case is the PROWESS clinical trial~\citep{Bernard-etal2001}, which randomized 1,690 patients with severe sepsis to receive the biologic drug Xigris. The treatment, administered intravenously, demonstrated a notable benefit--- a 6\% reduction in 28-day mortality compared to the 31\% mortality observed in the control group. This apparent effectiveness prompted early termination of the trial and led to the FDA’s accelerated approval of Xigris for severe sepsis in 2001.
However, controversy soon followed. A separate analysis indicated that the intervention was associated with a 1.5\% increase in the incidence of serious bleeding~\citep{Siegel2002}, raising concerns about the potential harm to patient safety. Ultimately, growing evidence of adverse effects led the manufacturer to voluntarily withdraw the drug from the global market in 2011~\citep{Food2011}.

In this article, we aim to estimate ITRs that maximize the reward while keeping the induced harm rate below a pre-specified threshold. Our main contributions are threefold. First, we theoretically derive the explicit form of the oracle ITR. We demonstrate that the harm rate threshold corresponds to a risk-aversion factor, and the oracle ITR is the indicator function of an adjusted CATE --- the CATE minus the risk-aversion factor multiplied by the conditional treatment harm rate (CTHR). We can view the adjusted CATE as a utility function, with the oracle ITR assigning treatment when the utility exceeds zero. This interpretation offers an intuitive understanding of the threshold and  highlights the tradeoff between maximizing reward and controlling harm. 
 {Second}, given that the CTHR is generally unidentifiable as it involves the joint distribution of potential outcomes, we propose two strategies for estimating ITRs when the CTHR is partially identifiable. 
The first strategy adopts a pessimistic approach by controlling the worst-case harm rate, achieved by substituting the CTHR with its sharp upper bound. 
 However, note that the upper bound is attainable only under extreme data-generating mechanisms, 
  the first strategy may be overly conservative in practice.  We thus introduce an improved strategy that incorporates expert knowledge of the correlation between potential outcomes. The key idea is to reformulate the CTHR using the conditional correlation coefficient between potential outcomes. 
  In real-world applications, domain experts may provide a plausible range for this coefficient~\citep{Bodik-etal2025}. We further demonstrate that, even in the absence of precise expert input, assuming the sign of the coefficient, e.g, a positive correlation,  can substantially tighten the upper bound, thereby resulting in less conservative ITRs.   

Third, we establish the large-sample properties of the proposed estimator of ITRs, proving its 
$L_1$-consistency and deriving its convergence rate of excess risk, defined as the average utility loss incurred when the optimal target ITR is replaced by its estimator. 
Moreover, we empirically evaluate the proposed methods through a simulation study. We further 
 apply the proposed method to analyze two real-world datasets from a new perspective,  estimating ITRs with controllable harm rates and assessing the extent to which potential harm can be mitigated relative to historical treatment decisions under different harm rate thresholds. 
There has been growing interest in estimating ITRs that not only maximize the reward but also adhere to key principles such as explainability~\citep{Dazeley-etal2023, Stephanie-etal2024}, fairness~\citep{Wang-etal2018-quantile, Fang-etal2023, Qi-etal2023}, and harmlessness or safety~\citep{Wang-etal2018}. Our method falls within this line of research and is closely related to the work of \citet{Wang-etal2018}.
     \citet{Wang-etal2018} proposed a method for estimating ITRs that balance reward and risk. However, their definition of risk refers to adverse events captured by a separate outcome---fundamentally different from our concept of harm, as we define both reward and harm using the same outcome. Moreover, the harm rate in this study involves the joint distribution of potential outcomes and may face partial identification issues, which \citet{Wang-etal2018} does not. 

The remainder of this paper is organized as follows. Section \ref{sec:setup-pre} describes the setup.  
  Section \ref{sec:optial-ITR} derives the oracle ITR with controllable harm rates. Section \ref{sec4-1-1} proposes a pessimistic strategy for estimating ITRs with controllable harm rates and establishes the large-sample properties. Section \ref{sec4-2-3} proposes an improved strategy by integrating expert knowledge. Section \ref{sec:simulation} conducts simulation studies to assess the performance of the proposed methods. Section \ref{sec:application} illustrates the proposed methods using two real-world datasets. Section \ref{sec:conclusion} concludes with a discussion.

\section{Setup}
\label{sec:setup-pre}

\subsection{Notation}

We adopt the potential outcomes framework~\citep{Rubin1974, Neyman1990} to define causal effects. 
Each individual is represented by a random tuple $\{X, A, Y(1), Y(0)\}$ drawn from a superpopulation $\mathbb{P}$, where $X \in \mathcal{X}$ denotes pre-treatment covariates, $A \in \mathcal{A} =\{0, 1\}$ is a binary treatment variable ($A = 1$ for treatment, $A = 0$ for control), and $Y(a) \in \{0, 1\}$ is the potential binary outcome had the individual received treatment $a$, for $a = 0, 1$.
We assume the {stable} unit treatment value assumption (SUTVA), meaning there are no multiple versions of the treatment and no interference between individuals. The observed outcome is then given by $Y = (1 - A)Y(0) + A Y(1)$. Without loss of generality, we define $Y = 1$ as a favorable outcome (e.g., survival) and $Y = 0$ as an unfavorable outcome (e.g., death). 
Suppose $\left\{\{X_i, A_i, Y_i(1), Y_i(0)\} : i = 1, \ldots, N\right\}$ is a simple random sample of $N$ individuals from the superpopulation $\mathbb{P}$, with the corresponding observed data denoted as $\{(X_i, A_i, Y_i) : i = 1, \ldots, N\}$.
The individual treatment effect is defined as $\tau_i = Y_i(1) - Y_i(0)$, representing the causal effect of the treatment on individual $i$. A positive $\tau_i$ indicates that the treatment benefits individual $i$, while a negative $\tau_i$ suggests it is harmful. However, since each individual receives only one treatment, only one of $\{Y_i(0), Y_i(1)\}$ is observed, making $\tau_i$ generally unidentifiable. In practice, the CATE is commonly used to guide individualized decisions \citep{Qian2011AoS, Zhao-etal2012,  Chen-etal2016}.  It is defined as $\tau(x) =\E\{Y(1) - Y(0)\mid X = x\},$ representing the average treatment effect within the subpopulation characterized by $X = x$. 
The CATE characterizes how the treatment effect varies across different subpopulations, thereby helping researchers identify which subpopulations may benefit from or be harmed by the treatment. 
 

  
However, as a subpopulation-level metric, $\tau(x)$ does not capture the inherent individual variability within each subpopulation, which may be important for individualized decision-making~\citep{Lei-Candes2021, 2022nathan, Wu-etal-2024-Harm}. For example, consider a study with two subpopulations, both with a CATE of 0.6. In the first subpopulation, the drug benefits 80\% of individuals $(Y(1) - Y(0) = 1)$ but harms the remaining 20\% $(Y(1) - Y(0) = -1)$. In the second subpopulation, the drug benefits 60\% of individuals while the rest remain unaffected $(Y(1) - Y(0) = 0)$. Although both subpopulations have identical CATEs, assigning treatment to the first subpopulation would result in harm to 20\% of individuals, while no one would be harmed in the second subpopulation. 

For quantifying the potential harm induced by treatment in the subpopulation with $X=x$,  we define the conditional treatment harm rate as  
$$\text{THR}(x) = \P(Y(0) = 1, Y(1) = 0 \mid X = x),$$  
which represents the proportion of individuals in the subpopulation for whom treatment is worse than control. Similarly, we define the conditional treatment benefit rate as 
		 $\text{TBR}(x) = \P( Y(0) = 0, Y(1) = 1 \mid X=x)$. 
Both $\text{THR}(x)$ and $\text{TBR}(x)$ provide complementary information to $\tau(x)$, capturing nuances that are essential for individualized decision-making. 			 	 

\subsection{Problem Description}  

Let $\pi(\cdot)$ be an ITR that maps from the covariate space $\mathcal{X}$ to the treatment space $\mathcal{A}$, representing a decision rule for whether an individual with covariates $X = x$ should receive treatment. For any given ITR $\pi$, we define its induced reward as: 
 $$\begin{gathered}
     R(\pi) = \E\left[\pi(X) Y(1) + \{1 - \pi(X)\}Y(0)\right] = \E\left\{ \pi(X) \tau(X)\right\} + \E\left\{ Y(0) \right\} ,  
 \end{gathered}$$   
which represents the average outcome when $\pi$ is applied to the entire population. In practice, treatment cost is also an important consideration. The reward can then be redefined as $R(\pi) = \E\left[\pi(X)\{Y(1)-c\} + {1 - \pi(X)}Y(0)\right]$, where $c$ denotes the cost of assigning treatment. 
For simplicity, we omit the cost term, as our main results naturally extend to this case. 
Similarly, the    treatment harm rate and   benefit rate induced by $\pi{(\cdot)}$ are defined as follows:  $$\begin{gathered}
 \text{THR}(\pi) =   \E\left\{  \text{THR}(X) \pi(X)\right\},\quad \text{TBR}(\pi) =  \E \left\{\text{TBR}(X)  \pi(X)\right\}.
 \end{gathered}$$  
  
In this paper, we aim to seek an ITR that maximizes the reward while keeping the induced THR below a threshold. Define the corresponding set of oracle ITRs as 
\begin{equation}  \label{eq1} 
  \Pi^*_\lambda :=   \begin{cases}    
        &  \arg\max_{\pi}  ~   R(\pi),   \\  
& s.t. ~  \text{THR}(\pi) \le \lambda, 
\end{cases}
\end{equation} 
where $\lambda$ is a pre-specified threshold representing the maximum acceptable level of THR. For example, $\lambda = 0.05$ limits the estimated ITR to harm fewer than 5\% of individuals. 
{The typical framework for estimating ITRs focuses solely on maximizing the reward $R(\pi)$~\citep{Wang-etal2018, Athey-Wager2021}. However, it does not ensure a controllable harm rate; see Remark \ref{Remark1} below for details.}  

\begin{remark}[Maximizing reward alone cannot control the harm rate]   \label{Remark1}
If we estimate the ITR by maximizing only the reward $R(\pi)$ while ignoring the potential harm it may induce, the oracle ITR is given by 
         \begin{align*} 
       {\pi_{\infty}^*(x) :=   \mathbb{I}\{\tau(x) > 0\} \in     \Pi^*_\infty := \arg \max_{\pi}R(\pi),}    
         \end{align*} 
   where $\mathbb{I}(\cdot)$ is the indicator function. The $\pi_\infty^*(x)$ assigns treatment for individuals whose $\tau(x)$ is greater than 0  and vice versa.   Note that $   \tau(x)  =  \text{TBR}(x)  - \text{THR}(x)$, 
            $\pi_\infty^*(x)$ may fail to control the harm rate effectively.  This is because $\tau(x)$ focuses solely on the difference between  $\text{TBR}(x)$  and  $\text{THR}(x)$, without explicitly controlling $\text{THR}(x)$ itself, potentially resulting in harmful treatment assignments for certain individuals. 
           For example, when both $\text{TBR}(x)$ and $\text{THR}(x)$ are large, a positive difference still lead to assigning treatment to the subpopulation with $X = x$, resulting in harmful decisions for a proportion of individuals. 
\end{remark}


\section{The Oracle ITR under Harm Constraint}
\label{sec:optial-ITR}

\subsection{The Oracle ITR under Harm Constraint}

In this section, we first derive the oracle ITR in the set $\Pi^*_\lambda$, assuming that the $\tau(X)$ and $\text{THR}(X)$ are known. This derivation motivates estimation methods proposed in Section \ref{est-FU}. 
  Intuitively, finding the oracle ITR in \eqref{eq1}  requires a careful balance between reward and harm. To formalize this trade-off, we categorize individuals into four groups based on the signs of $\tau(X)$ and $\text{THR}(X)$: 
 
\begin{itemize} \item {\( \mathcal{G}_1 \)}: This group satisfies $\tau(X) > 0$ and $\text{THR}(X) > 0$, indicating that while the treatment is beneficial on average, it may introduce some harm to certain individuals within this group.

\item {\( \mathcal{G}_2 \)}: This group satisfies $\tau(X) \leq 0$ and $\text{THR}(X) = 0$, indicating that the treatment is not beneficial on average and does not cause harm to any individuals.

\item {\( \mathcal{G}_3 \)}: This group satisfies $\tau(X) > 0$ and $\text{THR}(X) = 0$, indicating that the treatment is beneficial on average without causing harm to any individuals.

\item {\( \mathcal{G}_4 \)}: This group satisfies $\tau(X) \leq 0$ and $\text{THR}(X) > 0$, indicating that the treatment is not beneficial on average and would cause harm to certain individuals. \end{itemize}

From these categorizations, we should not assign treatment to individuals in groups $\mathcal{G}_2$ and $\mathcal{G}_4$, as doing so fails to increase reward and may even cause harm. In contrast, individuals in group $\mathcal{G}_3$ should assign treatment because it increases reward without causing harm. For individuals in group $\mathcal{G}_1$, the treatment decision involves a trade-off between reward and harm. Building on this insight, we derive a closed-form expression for the oracle ITR in $\Pi^*_\lambda$, as shown in Theorem \ref{thm1}.

\begin{theorem}[Oracle ITR] \label{thm1}
Given a pre-specified threshold $\lambda$,  

(a) If $\lambda < \E[\textup{THR}(X) \cdot \mathbb{I}\{\tau(X) > 0\}]$, the oracle ITR is obtained by
\begin{equation}
    \label{oracel-policy}\pi_{\lambda}^*(x) := \mathbb{I}\{\tau(x) - \beta^* \textup{THR}(x) > 0\},  
\end{equation}
where $\beta^*$ satisfies 
\begin{equation} \label{eq4} \E[ \textup{THR}(X) \cdot \I\{\tau(X) - \beta^* \cdot \textup{THR}(X) > 0\}] = \lambda. \end{equation}

(b) If $\lambda \geq \E[\textup{THR}(X) \cdot \mathbb{I}\{\tau(X) > 0\}]$, the oracle ITR reduces to
\[
\pi_\lambda^*(x) = \pi_\infty^*(x) = \mathbb{I}(\tau(x) > 0),
\]
which corresponds to the oracle ITR obtained by maximizing only the reward, without the constraint on harm rate.
\end{theorem}

As established in the proof, the parameter $\beta^*$ in \eqref{eq4} is essentially the Lagrange multiplier that arises when solving the constrained optimization problem via the Lagrangian approach. 
The optimal ITR $\pi_{\lambda}^*$ presented in Theorem~\ref{thm1} intuitively reflects the tradeoff between reward maximization and harm control. When the pre-specified threshold $\lambda$ is sufficiently large—specifically, when $\lambda$ is greater than or equal to 
 $$\E[\textup{THR}(X) \cdot \mathbb{I}\{\tau(X) > 0\}] =  \E\{  \textup{THR}(X)  \mid  X \in \mathcal{G}_1\}\cdot \P( X \in \mathcal{G}_1 ),$$
which is the product of the average treatment harm rate in group $\mathcal{G}_1$ and the proportion of $\mathcal{G}_1$ in the entire population,  
the constraint $  \text{THR}(\pi) \leq \lambda$ in \eqref{eq1}  {is automatically satisfied by maximizing only the reward,  leading to the unconstrained optimal ITR stated in Theorem~\ref{thm1}(b). 


From Theorem~\ref{thm1}(a), when \(\lambda\) is less than \(\E\left\{\textup{THR}(X) \cdot \mathbb{I}\{\tau(X) > 0\}\right\}\), the optimal ITR takes the form \(\mathbb{I}\{\tau(X) - \beta^* \cdot \text{THR}(X) > 0\}\). This decision rule assigns treatment only when \(\tau(X) = \text{TBR}(X) - \text{THR}(X)\) exceeds \(\beta^* \text{THR}(X)\), which is equivalent to requiring that the benefit rate be at least \((1 + \beta^*)\) times the harm rate.  Under this interpretation, we can view \(\tau(X) - \beta^* \text{THR}(X)\) as a utility function that captures the optimal tradeoff between benefit and harm rates, where larger values of \(\beta^*\) reflect greater caution in assigning treatment. Accordingly, we define the expected utility induced by an ITR $\pi$ as $ \E[\pi(X) \{\tau(X) - \beta^* \text{THR}(X)\}],$ 
where the expectation is taken with respect to the distribution of $X$. 
Furthermore, equation~\eqref{eq4} establishes a monotone relationship between $\beta^*$ and $\lambda$: as the harm rate threshold $\lambda$ decreases, the risk-aversion parameter $\beta^*$ increases correspondingly. 
 \subsection{Assumptions and Partial Identification} 

Throughout this section, we maintain the following strong ignorability assumption. 
  \begin{assumption} \label{assump1}
(i) $\{Y(0),Y(1)\} \indep  A \mid X$; (ii) $\epsilon  <  \P(A = 1 \mid X = x) < 1 - \epsilon$ for all $x\in \mathcal{X}$, 
where $0 < \epsilon < 1/2$ is a constant. 
\end{assumption} 
 
Assumption \ref{assump1} requires that $X$ includes all confounders affecting both the treatment and the outcome and it holds in either randomized controlled trials or unconfounded observational studies. 
Assumption \ref{assump1} is a standard assumption in causal inference~\citep{Imbens-Rubin2015, Hernan-Robins2020}. Under Assumption \ref{assump1},  the CATE is identified as $\tau(x) =  \mu_1(x) - \mu_0(x)$, where $\mu_a(X) = \E(Y \mid A =a, X = x)$.

However,  identifying $\text{THR}(x)$ generally requires additional assumptions, such as  monotonicity ($Y(1)\geq Y(0)$, \citeauthor{Huang-etal2012}, \citeyear{Huang-etal2012}) and conditional independence ($Y(0) \indep Y(1) \mid X$, \citeauthor{Shen-etal2013}, \citeyear{Shen-etal2013}). These assumptions are stringent and rarely satisfied in practice. 
Since $\text{THR}(x)$ depends on the joint distribution of potential outcomes, while only one outcome is observable for each individual, $\text{THR}(x)$ is generally unidentifiable, even in randomized controlled trials~\citep{MuellerPearl2023-CausalInference, Mueller-Pearl2023}. As a result, estimating the oracle ITR $\pi_{\lambda}^*(x)$ in \eqref{oracel-policy} is infeasible in practice.  
Typically, when an estimand is not identifiable, the focus shifts to partial identification, which entails determining its bounds. 

Under Assumption \ref{assump1}, previous studies~\citep{Zhang-etal2013, 2022nathan} have obtained the sharp Fréchet--Hoeffding lower and upper bounds on $\mathrm{THR}(x)$, given by:
\begin{equation}
    \label{eq:bounds-FH}
    \text{THR}_L(x) = \max \{ 0, -\tau(x) \}, \quad \text{THR}_U(x) = \min \{ \mu_0(x), 1 - \mu_1(x) \},
\end{equation}
which represent the best-case and worst-case harm rates for the subpopulation with $X = x$. 

Let $\mathcal{H(P)}$ represent the set of all compatible $\text{THR}(\cdot)$ functions, which are bounded by the functions $\text{THR}_L(\cdot)$ and $\text{THR}_U(\cdot)$. In the context of ITR estimation under partial identification, the minimax regret method is widely used for solving ITRs within $\mathcal{H(P)}$ or within its subset restricted by a specified function class~\citep[see e.g.,][]{Manski2011, Kallus2021, Pu-Zhang2021, Adamo2022, Yata2023, Eli-etal2023,cui2025policy}. 
However, we cannot directly apply a similar estimation method here. Although our optimization objective in \eqref{oracel-policy} similarly involves a partially identified quantity $\text{THR}(X)$, it differs in that it also includes an additional parameter $\beta^*$. The solution for $\beta^*$ depends on the estimation equation in \eqref{eq4}, which also involves the unidentifiable quantity $\text{THR}(X)$.

Besides the minimax approach, the original optimization problem in \eqref{eq1} shares similarities with constrained optimization problems that maximize benefit subject to certain constraints~\citep{2018Who,Carneiro2018,Sun2024}. However, our problem presents a distinct challenge: the constraint $\text{THR}(\pi) \le \lambda$ in \eqref{eq1} depends on harm rate that are not identifiable from observed data. As a result, existing methods may not be directly applicable to our setting. 

\section{A Pessimistic Strategy}    \label{sec4-1-1} 
As discussed previously, assumptions regarding the identifiability of $\text{THR}(X)$ may be difficult to justify in practice. In this section, we propose a simple yet conservative strategy for learning ITRs by controlling the worst-case harm rate. The key idea is to replace the unidentifiable constraint in \eqref{eq1} with its sharp upper bound. Specifically, using the Fréchet--Hoeffding upper bound $\text{THR}_U(x)$ from \eqref{eq:bounds-FH}, we define the pessimistic ITR set as: 
\begin{align}   \label{eq-6} 
   {\Pi^\dag_\lambda}   :=  \begin{cases}   
         & \arg\max_{\pi}  ~   R(\pi)   \\
& \text{s.t.} \quad   {\text{THR}}_U(\pi)  \le \lambda, 
\end{cases} 
\end{align}
where ${\text{THR}}_U(\pi) = \E\{ {\text{THR}}_U(X)  \pi(X)  \}$ is the sharp upper bound on $\text{THR}(\pi)$. This approach is more conservative than the oracle formulation in $\Pi^*_\lambda$ since it imposes the strictest feasible constraint on harm rate.

Analogous to Theorem \ref{thm1}, we can characterize the optimal solution:
$$\pi_\lambda^\dag(x) = \mathbb{I}\{\tau(x) - \beta^\dag \textup{THR}_U(x) > 0 \},$$ 
where the Lagrange multiplier $\beta^\dag$ satisfies
$$\E[ \text{THR}_U(X) \cdot \mathbb{I}\{\tau(X) - \beta^\dag \text{THR}_U(X) > 0\}] = \lambda.$$ 
Importantly, this formulation admits a max-min interpretation. The pessimistic problem \eqref{eq-6} is equivalent to:
\begin{gather*}
\arg\max_{\pi} \left\{ R(\pi) - \beta^\dag \textup{THR}_U(\pi) \right\} = \arg \max_{\pi} \min_{\text{THR}(x) \in \mathcal{H(P)}} \left\{ R(\pi) - \beta^\dag \textup{THR}(\pi) \right\}.   
\end{gather*}
which optimizes against the worst-case harm function within the identified set $\mathcal{H(P)}$. 
While $\pi_\lambda^\dag(x)$ may be overly conservative and suboptimal compared to the oracle policy $\pi_{\lambda}^*$ \citep{cui2021individualized}, it offers a crucial practical advantage: by construction, it guarantees that the true harm rate never exceeds $\lambda$, making it particularly valuable when safety is paramount.

In Section \ref{sec4-1-3} of the Supplementary Material, we provide a slightly compromised strategy to control the worst-case with high probability. To achieve this, we define \( \text{THR}_{\alpha} \) as the \( (1-\alpha) \)-th quantile of \( \text{THR}_U(X) \), i.e.,  $
\P( \text{THR}_U(X) \leq  \text{THR}_{\alpha} ) \geq 1 - \alpha.$ 
We then replace \( \text{THR}_U(x) \) with \( \text{THR}_{\alpha}(x) := \min\{ \text{THR}_{\alpha}, \text{THR}_U(x) \} \)
and use the same estimation procedures as the pessimistic strategy to estimate the corresponding ITRs. We omit the detailed discussion in the main text. 

\subsection{Estimation Procedure}
\label{est-FU}

    \begin{algorithm}[t]
\caption{Estimation of the target ITR {$\pi_\lambda^\dag(x)$}} \label{algorithm1}
\begin{algorithmic}[1]
\item[\textit{Step 1 (sample splitting).}] We randomly partition the data into $K$ distinct groups, each containing $N/K$ observations (for simplicity, assuming $N/K$ is an integer), and define $\mathcal{I}_{1}$ through $\mathcal{I}_{K}$ as the corresponding index sets. Additionally, let $\mathcal{I}_{k}^{c} = \{1, ...,  N\}\setminus \mathcal{I}_k$ denote the complement of $\mathcal{I}_{k}$ for $k = 1, \ldots, K$. 

\item[\textit{Step 2 (nuisance parameter training).}]  

 \medskip
  {for } $k = 1$ {to} $K$ {do} 
\begin{itemize}
    \item[(1)]  Estimate $\tau(x)$ and $\text{THR}(x)$ using the sample indexed by $\cI_k^c$. We denote $\hat \tau^{(-k)}(x)$ and {$\widehat{\text{THR}}_U^{(-k)}(x)$} as the corresponding estimators;    
 
    \item[(2)] Obtain the predicted values of $\hat \tau^{(-k)}(X_i)$ and {$\widehat{\text{THR}}_U^{(-k)}(x)$} for $i \in \cI_k$. 
 
    \item[(3)]  Estimate {$\beta^\dag$}  by solving equation \eqref{eq-est-beta} with the sample indexed by $\cI_k$. 
  Denote $\hat \beta^{(k)}$ as the estimator of {$\beta^\dag$} . 
\end{itemize} 
{end}

\medskip  \noindent
\item[\textit{Step 3 (estimation of the target ITR).}] 
For $i \in \mathcal{I}_k$ where $k = 1, ..., K$,  the estimator of the target ITR is given by ${\hat \pi_{\lambda}^\dag(X_i)} = \mathbb{I}(\hat \tau^{(-k)}(X_i) - \hat \beta^{(k)} \cdot {\widehat{\text{THR}}_U^{(-k)}(X_i)} > 0)$.  
\end{algorithmic}
\end{algorithm}

 The estimation method for {$\pi_\lambda^\dag(x)$}  are provided in Algorithm \ref{algorithm1}. 
 In Step 1, we adopt the sample-splitting technique to accommodate complex nonparametric estimators for the nuisance parameters $\tau(x)$ and {$\text{THR}_U(x)$}, thereby circumventing the need for empirical process conditions and simplifying the theoretical analysis \citep{Chernozhukov-etal-2018,wager2018estimation, Athey-Tibsirani-Wager-2019, Kennedy-2020}.  
 
  For Step 2, we can use the various methods for estimating $\tau(x)$ and $\text{THR}(x)$, such as the generalized linear model and sieve method.
  We use the grid search method for estimating {$\beta^\dag$} as it is intuitive and easy to implement. 
  Specifically, we estimate {$\beta^\dag$}  by solving the sample moment equation:     
    \begin{equation}  \label{eq-est-beta}
      \frac{1}{ |\cI_k| }  \sum_{i \in \cI_k } \widehat{\text{THR}}^{(-k)}_U(X_i) \cdot  \I\{\hat \tau^{(-k)}(X_i) - \beta \cdot \widehat{\text{THR}}^{(-k)}_U(X_i) > 0\} - \lambda = 0.    
    \end{equation}

After obtaining the estimator of {$\beta^\dag$} , if we do not impose any restriction on the function class of ITRs, it is natural to estimate $\pi_\lambda^\dag(x)$ with the plug-in method, as shown in Step 3.  Similarly, for a new test dataset indexed by $\mathcal{I}_{test}$ that contains only covariates, we can use all observed data $\{(X_i, A_i, Y_i): i = 1, \ldots, N\}$ to estimate $\tau(x)$ and {$\text{THR}_U(x)$}, obtain the predicted values of $\tau(X_i)$ and {$\text{THR}_U(X_i)$} for $i \in \mathcal{I}_{test}$,  estimate {$\beta^\dag$}  using equation \eqref{eq-est-beta} by replacing $\mathcal{I}_k$ with $\mathcal{I}_{test}$, and estimate $\pi_\lambda^\dag(x)$ in the test dataset using the plug-in method.

\subsection{Theoretical Analysis} 
\label{sec:theoretical}
We present the large sample properties of the proposed estimator {$\hat \pi^\dag_\lambda(x)$}.
Similar to the discussion in the paragraph below Theorem \ref{thm1}, for the optimization problem \eqref{eq-6}, when {$\hat \pi_\lambda^\dag(x)$} is applied to the entire population,   
the induced expected utility is defined as 
 {$U(\hat \pi_\lambda^\dag) = \E[ \hat \pi_{\lambda}^\dag(X) \{\tau(X) - \beta^\dag  \text{THR}_U(X)\}]$.}  
Note that $U(\hat \pi_\lambda)$  is a random variable, 
as the expectation is taken with respect to the distribution of $X$. 
The difference between this utility and the optimal utility  {$U(\pi_\lambda^\dag)$} is the risk bound (or excess risk), defined by {$U(\pi_\lambda^\dag) - U(\hat \pi_\lambda^\dag)$}, which measures the utility loss incurred by replacing the optimal ITR $\pi_{\lambda}^\dag(x)$ with its estimator $\hat \pi_{\lambda}^\dag(x)$.    
 
First, we discuss the consistency of {$\E\{ |\hat \pi_\lambda^\dag(X) - \pi_\lambda^\dag(X) | \}$} and {$U(\pi_\lambda^\dag) - U(\hat \pi_\lambda^\dag)$}, which relies on Conditions \ref{cond1} and \ref{cond2} defiend below. Let {$\bm{\eta}(x) := \{\tau(x), \text{THR}_U(x) \}$} be the nuisance parameters and  
$\Psi(\beta; \bm{\eta}) := {\E[ \textup{THR}_U(X) \cdot  \I\{\tau(X) - \beta \cdot \textup{THR}_U(X) > 0\}]} - \lambda$, then the population estimating equation for {$\beta^\dag$} is  $\Psi(\beta; \bm{\eta}) = 0$.

\begin{condition} \label{cond1} 
 (i) $\Psi(\beta; \bm{\eta})$ is continuously differentiable with respect to $\beta$ at a neighborhood of {$\beta^\dag$}; 
 (ii)   $\beta \in [0, M]$ for a finite constant $M$ and 
 {$\beta^\dag$} is an interior point of $[0, M]$; and 
 (iii) $|\partial \Psi(\beta^\dag, \bm{\eta})/\partial \beta | \geq c > 0$ for a constant $c$.
\end{condition}
  
Condition \ref{cond1} states the regularity conditions for $\Psi(\beta; \bm{\eta})$.  Given that $\Psi(\beta; \bm{\eta})$ is a monotone function of $\beta$, Condition \ref{cond1} ensures that $\beta^\dag$ is the unique solution to $\Psi(\beta; \bm{\eta}) = 0$. It is noteworthy that although Condition \ref{cond1} imposes a certain degree of smoothness on $\Psi(\beta; \bm{\eta})$ with respect to $\beta$, its sample counterpart (i.e., the left-hand side of equation $\eqref{eq-est-beta}$) remains a non-smooth function of $\beta$. Consequently, the standard Z-estimator theory (see e.g., Chapter 5 of \citet{vdv-1998}) cannot be directly applied to establish the consistency of $\hat{\beta}^{(k)}$. To address this, we further introduce Condition \ref{cond2}.   


\begin{condition} 
 \label{cond2}
The density of the random variable $\tau(X) - \beta  \cdot{\textup{THR}_U(X)}$ is bounded for any $\beta$.   
\end{condition} 

Condition \ref{cond2} is mild given that both $\tau(X)$ and {$\textup{THR}_U(X)$} are bounded functions for binary outcomes.  Essentially, Condition \ref{cond2}   restricts the behavior of $\tau(X) - \beta \cdot \text{THR}_U(X)$ in the vicinity of the level $\tau(X) - \beta  \cdot\text{THR}_U(X) = 0$, which is crucial for addressing the discontinuity of the sample estimating equation \eqref{eq-est-beta}. Without loss of generality, suppose that the density function of $\tau(X) - \beta  \cdot\textup{THR}_U(X)$ is bounded by a constant $\delta$, then for any $t > 0$,  Condition \ref{cond2}  implies that 
  $\P( | \tau(X) -  \beta \cdot \text{THR}_U(X) | \leq t ) \leq 2 \delta t$. This is a specific case of the margin condition ($\P\big (  | \tau(X) - \beta \cdot \textup{THR}_U(X) |  \leq t  ) \leq C_0  t^{\gamma}$ for some constants $C_0$ and $\gamma > 0$), which is commonly employed in the estimation of non-smooth functions~\citep{Audibert-etal-2007, Luedtke-etal-2016, 2018Who, Kennedy-2019}. See Lemma S2 and  Condition S1 in Supplementary Material for more details.  
   

\begin{lemma} \label{lem1} Under Assumption \ref{assump1},  Conditions \ref{cond1} and \ref{cond2}, if $\hat \tau^{(-k)}(x)$ and $\widehat{\textup{THR}}_U^{(-k)}(x)$ are uniformly consistent estimators of $\tau(x)$ and $\textup{THR}_U(x)$, respectively, then  $\hat \beta^{(k)} \xrightarrow{\P} \beta^\dag$.  
\end{lemma}

Lemma \ref{lem1} establishes the consistency of  $\hat{\beta}^{(k)}$, which lays the foundation for 
the consistency of  {$\E\{ |\hat \pi_\lambda^\dag(X) - \pi_\lambda^\dag(X) | \}$  and $U(\pi_\lambda^\dag) - U(\hat \pi_\lambda^\dag)$}.   

\begin{theorem}[Consistency]\label{thm2} Under the same conditions of Lemma \ref{lem1}, we have that   
(a) $\E\{ |\hat \pi_\lambda^\dag(X) - \pi_\lambda^\dag(X) | \} = o_{\P}(1)$; 
(b) $U(\pi^\dag_\lambda) - U(\hat \pi_\lambda^\dag)= o_{\P}(1)$.    
\end{theorem}

Theorem \ref{thm2}(a) indicates that $\hat \pi_\lambda^\dag(x)$ converges to $\pi_\lambda^\dag(x)$ in terms of $L_1$-norm. However, pointwise convergence $\hat \pi_\lambda^\dag(x) \xrightarrow{\P} \pi_\lambda^\dag(x)$ does not hold, as we cannot guarantee that $\hat \tau^{(-k)}(x) - \hat \beta^{(k)} \cdot \widehat{\text{THR}}_U^{(-k)}(x)$ and  $\tau(x) - \beta^\dag \cdot \text{THR}_U(x)$ always have the same sign. Nevertheless, Condition \ref{cond2} ensures that the probability of such sign discrepancies is small, leading to the result in Theorem \ref{thm2}(a). Theorem \ref{thm2}(b) shows that the utility induced by $\hat \pi_\lambda^\dag(x)$ converges to the true optimal utility.  Next, we explore the convergence rate of the risk bound $U(\pi^\dag_\lambda) - U(\hat \pi_\lambda^\dag)$. 

%



\begin{condition} \label{cond3}  We introduce the following rate constraints: 
      $\max_{1\leq k\leq K} || \hat \tau^{(-k)}(x) - \tau(x) ||_{\infty} = O_{\P}(a_N)$ and   
  $\max_{1\leq k\leq K} || \widehat{\textup{THR}}^{(-k)}_U(x) - \textup{THR}_U(x) ||_{\infty} = O_{\P}(r_N)$, where both $a_N$ and $r_N$ are positive sequence converging to zero as $N \to \infty$. 
\end{condition}

Condition \ref{cond3} is a high-level assumption regarding the convergence rates of the estimators for the nuisance parameters. For $x \in \mathbb{R}^d$, suppose the nuisance parameters belong to the  $(s, L, \mathbb{R}^d)$-H\"{o}lder class. Then, if we estimate them using appropriately selected local polynomial estimators, the uniform convergence rate is   $(\log N / N)^{-s/(2s +d) }$~\citep{Stone1982}. 

\begin{theorem} \label{thm3}  Under Assumption \ref{assump1},  Conditions \ref{cond1}, \ref{cond2}, and \ref{cond3}, we have 
    \[     U(\pi^\dag_\lambda) - U(\hat \pi_\lambda^\dag) \leq  O_{\P}(a_N^2 \vee r_N^2 \vee N^{-1})   \]
 where $a \vee r = \max\{a,  r\}$.  
\end{theorem}

Theorem \ref{thm3} presents the upper bound for $U(\pi^\dag_\lambda) - U(\hat \pi_\lambda^\dag)$, with a convergence rate of order $O_{\P}(  a_N^2 \vee r_N^2 \vee N^{-1})$, which is faster than that of the nuisance parameter estimators. This result is not surprising, as Condition \ref{cond2} is a specific case of the margin condition, which is often used to accelerate the convergence rate of plug-in classifiers~\citep{Audibert-etal-2007} and parameters in non-smooth estimating equations~\citep{Bonvini-Kennedy2022}. 
     




\section{Improved Strategy by Integrating Expert Knowledge} \label{sec4-2-3}
The estimation methods for ITRs in Section \ref{sec4-1-1}  rely on the Fréchet--Hoeffding upper bound on $\text{THR}(x)$. Although it is wide and is attained only under extreme data-generating processes, this upper bound is sharp and we cannot improve it without introducing additional assumptions. In this subsection, we propose a practical and user-friendly method to improve the upper bound on $\text{THR}(x)$ by incorporating possible expert knowledge about the correlation coefficient of potential outcomes, thereby proposing a less conservative method for estimating ITRs with a controllable harm rate.  

The proposed method builds on an equivalence between $\text{THR}(X)$ and the Pearson  correlation coefficient $\rho(X) = \text{Corr}(Y(0), Y(1) \mid X)\in[-1,1]$, as explored by \citet{Wu-etal-2024-Harm}.
Concretely, under Assumption \ref{assump1}, we have 
    \[ \begin{aligned}
        \rho(X) &= \frac{ \E\{Y(0) Y(1)\mid X\}- \E\{Y(0)\mid X)\}  \E\{Y(1)\mid X)\} } {\sqrt{ \V\{Y(0)\mid X)\} \cdot  \V\{Y(1)\mid X)\}}}  = \frac{\mu_0(X)\{1-\mu_1(X)\} - \text{THR}(X) }{  \sqrt{  \mu_0(X)\{1-\mu_0(X)\} \mu_1(X)\{1-\mu_1(X)\}   } },  
    \end{aligned}  \] 
  where the second equality follows from the fact that, under  Assumption \ref{assump1},  $\E\{ Y(0) Y(1) \mid X\} = \P(Y(0) =1, Y(1)=1 \mid X) = \P(Y(0)=1\mid X) - \text{THR}(X)$ and $ \V(Y(a) \mid X) = \mu_a(X)(1 - \mu_a(X))$ for $a = 0, 1$.  In real-world applications, domain experts may provide a plausible range for, or at least the sign of $\rho(x)$, see Remark \ref{rmk2} below. 
  
   \begin{remark}  \label{rmk2}
\citet{Bodik-etal2025} demonstrate the plausibility of $\rho(x)$ across various common settings and through extensive simulations. The authors state in the abstract: 
   ``... although $\rho$ is fundamentally unidentifiable, we argue that in most real-world applications, it is possible to impose reasonable and interpretable bounds informed by domain-expert knowledge". 
\end{remark}

Once the correlation $\rho(X)$ is known, $\text{THR}(x)$ can be uniquely determined as:
\begin{equation}  \label{eq9}
    \text{THR}(x) =     \mu_0(x)\{1 - \mu_1(x)\} -    \rho(x) \sqrt{  \mu_0(x)\{1-\mu_0(x)\} \mu_1(x)\{1-\mu_1(x)\}    } . 
\end{equation}    
From equation \eqref{eq9}, $\text{THR}(x)$ is a decreasing function of $\rho(x)$. 
Therefore, we consider obtaining the upper bound on $\text{THR}(x)$ by determining the lower bound for $\rho(x)$, as in practice, we may have stronger prior information about $\rho(x)$. 
With known marginal distributions of the potential outcomes (restricted by  Assumption \ref{assump1}), the range of $\rho(x)$ is not necessarily the full interval $[-1, 1]$. The following Lemma \ref{lem2} summarizes the possible range of $\rho(x)$ under Assumption \ref{assump1}. 

\begin{lemma} \label{lem2} 
    Under Assumption \ref{assump1}, the lower bound $\rho_L(x)$ and upper bound $\rho_U(x)$ of $\rho(x)$ can be identified as:
    \begin{align*} 
    \rho_L(x) ={}& - \frac{ \min\big\{\{1-\mu_0(x)\} \{1-\mu_1(x) \}, \mu_0(x)\mu_1(x)\big\} }{\sqrt{ \mu_0(x)\left\{1-\mu_0(x)\right\} \mu_1(x)\left\{1-\mu_1(x)\right\} }}, \\
    \rho_U(x) ={}& \frac{ \min\big\{\mu_0(x) \{1-\mu_1(x) \}, \mu_1(x) \{1-\mu_0(x) \} \big\} }{\sqrt{ \mu_0(x) \{1-\mu_0(x) \} \mu_1(x)\left\{1-\mu_1(x)\right\} }}.
    \end{align*} 
\end{lemma}

When  $\rho(x) = \rho_L(x)$, equation \eqref{eq9} reduces to the Fr\'{e}chet--Hoeffding upper bound in \eqref{eq:bounds-FH}.  Since this upper bound is rarely attained in practice, $\rho_L(x)$ serves as a highly conservative lower bound for $\rho(x)$. It is important to note that reformulating $\text{THR}(x)$ in terms of $\rho(x)$ offers an opportunity to incorporate expert knowledge. 
Without loss of generality, we assume that $\tilde{\rho}_L(x)$ represents a pre-specified lower bound for the correlation coefficient, as provided by experts. 

\begin{assumption}[Expert Knowledge] \label{assump3}
     $\rho(x) \geq \tilde \rho_L(x)$. 
\end{assumption}

When $\tilde{\rho}_L(x) \leq \rho_L(x)$ for certain $x$, the expert knowledge in Assumption \ref{assump3} provides no additional information about $\text{THR}(x)$. However, if $\tilde{\rho}_L(x) > \rho_L(x)$ for certain $x$, we can use expert knowledge to  tighten the upper bound on $\text{THR}(x)$.  

\begin{proposition}[Improved Upper Bound on $\text{THR}(x)$] \label{prop2} Let  $\bar \rho(x) := \max\{\tilde \rho_L(x), \rho_L(x)\}$, the sharp upper bound on $\textup{THR}(x)$ under Assumptions \ref{assump1} and \ref{assump3} is  
  \[  \bar{\textup{THR}}_U(x):= \max \Big\{  \mu_0(x) \{ 1 - \mu_1(x) \} -    \bar \rho(x)  \sqrt{  \mu_0(x)\{1-\mu_0(x) \} \mu_1(x)\{1-\mu_1(x) \}  },  0  \Big\}, \]
which is not greater than the Fr\'{e}chet--Hoeffding upper bound under Assumption  \ref{assump1}.  
\end{proposition} 

Proposition \ref{prop2} indicates that incorporating expert  knowledge on the lower bound of $\rho(x)$ provides potential benefit for improving the upper bound on $\text{THR}(x)$. 
In real-world applications, we may not be able to specify the values of $\tilde{\rho}_L(x)$ due to limited expert knowledge. In such cases, we can instead consider qualitative information, such as the sign of $\rho(x)$. For example, assuming  $\tilde{\rho}_L(x) = 0$ in Assumption \ref{assump3} implies that $Y(0)$ and $Y(1)$ are positively correlated given $X = x$. 
  The positive correlation assumption is generally mild in practice~\citep{Bodik-etal2025}. 
   For example, in medical studies,  let $Y$ represent the disease status, a patient's health status (unmeasured factors)  affects both $Y(0)$ and $Y(1)$ and leads to a positive correlation between them~\citep{efron1991compliance}.
Note that $\rho_L(x)$ defined in Lemma \ref{lem2} is always less than or equal to zero. If we set $\tilde \rho_L(x) \geq 0$ in Assumption \ref{assump3}, we can tighten the upper bound on $\text{THR}(x)$ by ruling out possible negative values of $\rho(x)$. Formally, under Assumptions \ref{assump1} and \ref{assump3} with $\tilde \rho_L(x) \geq 0$, 
  \begin{align*}
   \text{THR}(x) 
          {}&\leq \max \Big\{  \mu_0(x) \{1 - \mu_1(x)\} -   \tilde \rho_L(x) \sqrt{  \mu_0(x)\{1-\mu_0(x)\}\mu_1(x)\{1-\mu_1(X)\}    },  0  \Big\} \\ 
      \leq{}&   \mu_0(x) \{1 - \mu_1(x)\}  \\
      \leq{}&  \min\{\mu_0(x), 1- \mu_1(x) \}\qquad \text{(Fr\'{e}chet--Hoeffding upper bound under Assumption \ref{assump1})}.
 \end{align*} 
In this case, we obtain a tighter upper bound on $\text{THR}(x)$ that may be substantially smaller than the  Fr\'{e}chet--Hoeffding upper bound under Assumption \ref{assump1}. 

Similar to \eqref{eq-6}, we define the set of improved target ITRs as: 
\begin{align*}    
 \Pi^\ddag_\lambda   :=   \begin{cases}  
       &   \arg\max_{\pi}  ~   R(\pi),  ~~  \\
& s.t. ~  \bar{\mathrm{THR}}_U(\pi)   \le \lambda, 
    \end{cases}
\end{align*}
where $\bar{\mathrm{THR}}_U(\pi) =  \E\{\bar{\mathrm{THR}}_U(X)\pi(X)\}$.  
Using the estimation procedure in Section \ref{est-FU}, we can solve for the ITRs by replacing the Fréchet--Hoeffding upper bound with the improved bound. Furthermore, similar asymptotic results, as discussed in Section \ref{sec:theoretical}, can also be derived.  
\section{Simulation Studies}
\label{sec:simulation} 

\subsection{Simulation Setup}

We conduct simulation studies to evaluate the finite-sample performance of the proposed ITR estimation methods. In this simulation, the baseline covariates  $ X = (X_1, X_2, X_3)^\intercal$ are drawn from a multivariate normal distribution  $\mathcal{N}(0, I_3)$, where $I_3$ is the $3\times 3$ identity matrix.  The treatment variable $A$ is assigned based on the probability $ \P(A=1 \mid X) = \Phi( (X_1 - X_2  +  X_3)/3)$, where $ \Phi(\cdot) $ denotes the cumulative distribution function of $\mathcal{N}(0, 1)$. 
We consider the following data-generating mechanism for potential outcomes $(Y(0), Y(1))$:  
    \begin{equation} \label{eq9-simu} \begin{split}
             \P(Y(0) = 1 \mid X, U) ={}& \Phi( (X_1 +  X_2 + X_3)/3 + \delta  U), \\
        \P(Y(1) = 1 \mid X, U) ={}& \Phi(  (X_1 +  X_2 + X_3)/3 + \delta  U/2  + 1/5),
    \end{split} 
    \end{equation}   
where $U$ is an unobserved variable drawn from $\mathcal{N}(0, 1)$. The observed outcome is $Y = A Y(1) + (1 - A) Y(0)$. The sample size is set to 1,000.
   
   In equation \eqref{eq9-simu}, the parameter $\delta$ controls the strength of unmeasured confounding between $Y(0)$ and $Y(1)$. 
   When $\delta = 0$, $\rho(x) \equiv 0$; when $\delta \neq 0$, the presence of $U$ induces a positive correlation between $Y(0)$ and $Y(1)$ given $X$, i.e., $\rho(x) > 0$. 
   To evaluate the robustness of the proposed methods, we consider three different values for $\delta$ (0, 1, and 2), and set the threshold $\lambda$ in the problem \eqref{eq1} to 0.02 and 0.05, respectively.


We compare the following five methods:
(a) CATE-Based: maximizing only the reward while ignoring the harm rate, $\hat \pi_{\infty}^*(x) = \mathbb{I}\{\hat \tau(x) > 0\}$ is the estimator of $\pi_{\infty}^*(x)$ defined in Remark \ref{Remark1}.  
(b) Pessimistic strategy: control the worst-case harm rate, see Section \ref{sec4-1-1} for details; (c) Conservative strategy: control the worst-case harm rate with a high probability of $0.90$, see Section \ref{sec4-1-3}  of the Supplementary Material  for details; (d) Expert knowledge ($\rho(x)\geq 0$):  incorporating expert knowledge that $\rho(x) \geq 0$, see Section \ref{sec4-2-3} for details; 
    (e) Expert knowledge  ($\rho(x)\geq 0.1$):  incorporating expert knowledge that $\rho(x) \geq 0.1$. 
For implementation, we use Probit regression to estimate $\mu_0(x)$ and $\mu_1(x)$, 
 which are then used to estimate $\tau(x)$, $\text{THR}_U(x)$, $\text{THR}_{\alpha}(x)$, and the improved upper bound on $\text{THR}(x)$ given in Proposition \ref{prop2}, by plugging in the estimates of $\mu_0(x)$ and $\mu_1(x)$. 
 



      





\subsection{Simulation Results}

We evaluate the performance of the comparison methods using two metrics: 
 (a) the treatment harm rate induced by the estimated ITR, defined by $\widehat{\text{THR}}(\hat \pi) := N^{-1} \sum_{i=1}^N \hat \pi(X_i) \cdot \mathbb{I}( Y_i(1) -Y_i(0) <0   )$, where $\hat{\pi}$ denotes a general estimated ITR; (b) the reward induced by the estimated ITR, defined by $\hat{R}(\hat \pi) := N^{-1} \sum_{i=1}^N \hat \pi(X_i)  Y_i(1)  + (1- \hat \pi(X_i))Y_i(0)$.
Each simulation is replicated 1,000 times. Figure \ref{fig1} presents boxplots of the harm rate and reward induced by the estimated ITRs for different methods, using a harm rate threshold of $\lambda = 0.05$. 
From Figure \ref{fig1}, we have the following observations.
 

\begin{figure}[!h]
\centering
\subfloat[Threshold $\lambda = 0.05$]{
\begin{minipage}[t]{0.5 \linewidth} 
\centering
\includegraphics[width=1\textwidth]{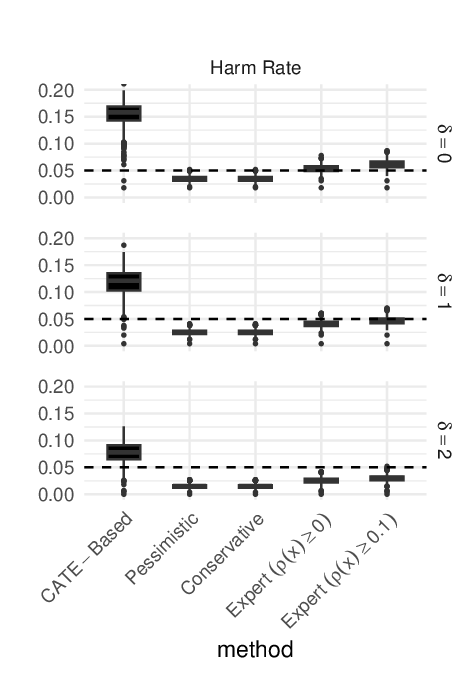}
\end{minipage}%
} \hspace{-20pt} 
\subfloat[Threshold $\lambda = 0.05$]{
\begin{minipage}[t]{0.5 \linewidth}
\centering
\includegraphics[width=1\textwidth]{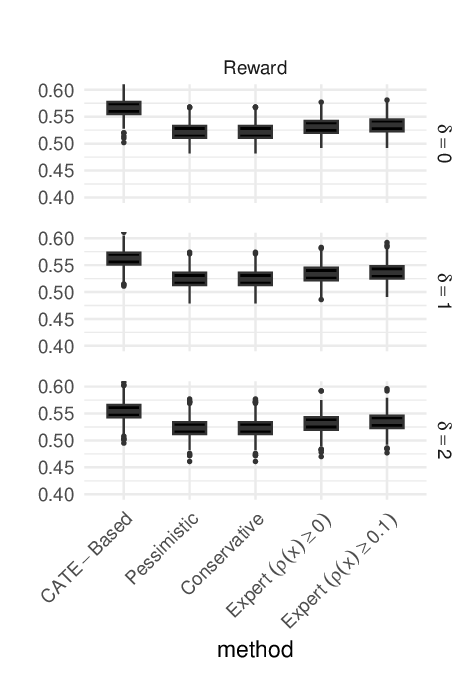}
\end{minipage} 
} 
\caption{(a) Boxplots of the harm rate induced by the estimated ITRs across different methods, where the dotted line indicating the harm rate threshold ($\lambda = 0.05$); (b) Boxplots of the reward induced by the estimated ITR across different methods.}  
\label{fig1} 
\end{figure}


First, there is a monotonic relationship between harm rates and rewards across all methods---lower harm rates correspond to higher rewards,  confirming their intrinsic tradeoff.  
In addition,  the five competing methods demonstrate a strict ordering in terms of harm rate: {CATE-Based $>$ Expert ($\rho(x)\geq 0.1$) $>$  Expert ($\rho(x)\geq 0$)  $>$ Conservative $\geq$ Pessimistic}, reflecting the strength of each method in controlling the harm rate.

Second, across all values of $\delta$, 
 the CATE-Based method exhibits a harm rate significantly exceeding the 0.05 threshold. In contrast, the Pessimistic, Conservative, and {Expert ($\rho(x)\geq 0$)} methods effectively maintain the harm rate near or below the 0.05 threshold.  
When $\delta = 0$,  the method incorporating expert knowledge $\rho(x) \geq 0.10$ (i.e., {Expert ($\rho(x)\geq 0.1$)}) yields  a harm rate slightly above 0.05. This is attributable to inaccuracies in the expert knowledge, as $\delta = 0$ implies $\rho(x) = 0$.
The  {Expert ($\rho(x)\geq 0$)} method maintains a harm rate close to 0.05 while achieving a relatively high reward, indicating its effectiveness when expert knowledge is valid in practice.  
  In terms of reward, the CATE-Based method achieves the highest value. However, the proposed methods (the remaining four approaches) markedly reduce the harm rate---from 0.15 to 0.05---while incurring only a minor tradeoff in reward, decreasing slightly from 0.56 to 0.52.  
When $\delta = 1$ or $2$, the expert knowledge offers additional information on the harm rate as discussed in Section \ref{sec4-2-3}, and the method leveraging this knowledge achieves the most competitive performance, yielding a higher reward while keeping the harm rate below 0.05.  
These observations demonstrate that the proposed methods are highly effective in controlling harm rates.

We present results using a harm rate threshold of $\lambda = 0.02$, which shows patterns similar to those in Figure \ref{fig1}. In addition, we report results using different cross-fitting folds, which demonstrate that the proposed methods perform stably across different folds. 
To avoid redundancy, we relegate them to Supplementary Material. 
\section{Application} 
\label{sec:application}

In this section, we demonstrate the proposed method through \emph{two real-world datasets} below. 
 
 \begin{itemize} \item The dataset from the Study to Understand Prognoses and Preferences for Outcomes and Risks of Treatments (SUPPORT)~\citep{connors1996effectiveness}.

 \item  The dataset from a large vocational training program called \emph{J\'{o}venes en Acci\'{o}n}~\citep{Attanasio-etal2017}. 
 \end{itemize}
\noindent 
In addition to the five methods compared in the simulation, we include a Naive ITR, denoted as $\hat{\pi}_{\text{naive}}(X) \equiv A$, for comparison. This serves as a baseline representing historical interventions. 
We use Probit regression to estimate $\mu_0(x)$ and $\mu_1(x)$.  
Let $\hat \mu_a(x)$ be the estimate of $\mu_a(x)$ for $a= 0, 1$.  Different from the simulation, we define the reward function induced by an estimated ITR $\hat{\pi}$ as 
$\hat{R}(\hat \pi) := N^{-1} \sum_{i=1}^N \hat \pi(X_i)  \hat \mu_1(X_i)  + (1- \hat \pi(X_i)) \hat \mu_0(X_i)$.  
Since the true harm rate is unknown in real-world datasets, we adopt the following three approaches to evaluate the harm rates induced by estimated ITRs: 
\begin{itemize}
   \item $\widehat{\text{THR}}_1(\hat{\pi}) := N^{-1} \sum_{i=1}^N \hat{\pi}(X_i) \cdot \widehat{\textup{THR}}_U(X_i)$, where $\widehat{\textup{THR}}_U(x) = \min\left\{\hat{\mu}_0(x), 1 - \hat{\mu}_1(x)\right\}$ is an estimator of the upper bound on  $\text{THR}(x)$ under Assumption \ref{assump1} only;

  \item $\widehat{\text{THR}}_2(\hat \pi) := N^{-1} \sum_{i=1}^N \hat \pi(X_i) \cdot  \widehat{\textup{THR}}_{U,\rho(x)\geq 0}(X_i)$, {where $\widehat{\textup{THR}}_{U,\rho(x)\geq 0}(x) = \hat \mu_0(x) (1- \hat \mu_1(x))$ is an estimator of the upper bound on $\text{THR}(x)$ under Assumption \ref{assump1} and expert knowledge of $\rho(x) \geq 0$;} 
  
\item $\widehat{\text{THR}}_3(\hat{\pi}) := N^{-1} \sum_{i=1}^N \hat{\pi}(X_i) \cdot \widehat{\textup{THR}}_{U,\rho(x)\geq 0.1}(X_i)$,   
where ${\widehat{\textup{THR}}_{U,\rho(x)\geq 0.1}(x)} = \max\{\hat \mu_0(x)(1- \hat \mu_1(x)) - 0.1 \sqrt{ \hat \mu_0(x)\hat \mu_1(x)(1- \hat \mu_0(x))(1- \hat \mu_1(x)) }, 0 \}$
is an estimator of the upper bound on $\text{THR}(x)$ under Assumption \ref{assump1} and expert knowledge that $\rho(x) \geq 0.1$.  
\end{itemize}
The three metrics essentially reflect varying levels of confidence in the harm rate estimates---ranging from the most pessimistic, $\widehat{\text{THR}}_1(\hat{\pi})$ (corresponds to $\rho(x) = \rho_L(x)$ in Lemma \ref{lem2}), to $\widehat{\text{THR}}_2(\hat \pi)$  (assuming  $\rho(x) \geq 0$), to $\widehat{\text{THR}}_3(\hat{\pi})$ (assuming $\rho(x) \geq 0.10$). These metrics serve as a form of sensitivity analysis for evaluating the harm rates induced by $\hat \pi$.  
We also use $\hat \E(\hat \pi) := N^{-1} \sum_{i=1}^N \hat \pi(X_i)$ to estimate the proportion of patients assigned treatment. 
For each method, we report point estimates of various evaluation metrics along with their 95\% confidence intervals (CIs), obtained via 1,000 nonparametric bootstrap replications.

\subsection{Safe ITRs in the SUPPORT}

{\bf Background.} The SUPPORT dataset  includes information on 5,735 critically ill patients in ICUs, 
 of whom 2,184 received Right Heart Catheterization (RHC) within 24 hours of admission (\(A = 1\)), while 3,551 patients served as the control group (\(A = 0\)). The outcome variable \(Y\) indicates whether survival time exceeded 30 days, with 3,817 patients surviving beyond 30 days and 1,918 patients passing away within this period. The baseline covariates \(X\) include 50 variables covering diagnostic details, comorbidities, vital signs, physiological measures, and demographic characteristics such as gender, race, age, education, and income.   
A controversial study by \citet{connors1996effectiveness} raised concerns that RHC may be ineffective and could even potentially cause harm to many patients. In response, several subsequent studies have assessed the impact of RHC on survival by estimating average treatment effects, such as \citet{Hirano-Imbens2001}, \citet{Tan2006}, \citet{Crump-etal2009} and  \citet{Vansteelandt-etal2012}.
\emph{We revisit the dataset from a different perspective, aiming to estimate ITRs with controllable harm rates and to evaluate the extent to which potential harm can be mitigated relative to historical treatment decisions.} 



\smallskip 
\noindent 
{\bf Results.} 
{The corresponding results are shown in Figures~\ref{fig2} and \ref{fig3}, with \(\lambda = 0.02\) and \(\lambda = 0.05\) as the respective tolerable harm rate thresholds.}  
Figure~\ref{fig2}(a) shows that the Naive method yields a substantially lower reward compared to the other methods. While the CATE-Based method achieves the highest reward, its advantage is not statistically significant relative to the {Expert ($\rho(x)\geq 0$) and Expert ($\rho(x)\geq 0.1$)}  methods. 
Figure~\ref{fig2}(b) shows that the Naive, Pessimistic, and Conservative methods assign a similar proportion of patients (38\%--40\%) to receive RHC, suggesting that historical interventions (as represented by the Naive method) were relatively conservative. In contrast, the aggressive CATE-Based method assigns a significantly higher proportion of patients (75\%) to receive RHC compared to the other approaches.  
Figures~\ref{fig2}(c)–\ref{fig2}(e) present the harm rates induced by the estimated ITRs. The Naive method yields a harm rate ranging from 0.055 to 0.079, slightly exceeding the threshold of 0.05, while the CATE-Based method results in a harm rate significantly above this threshold. In contrast, the Pessimistic, Conservative, and {Expert ($\rho(x)\geq 0$)} 
methods control the harm rate more effectively. In summary, when $\lambda = 0.05$, the Naive, Pessimistic, and Conservative methods assign similar proportions of patients to receive RHC. However, the Pessimistic and Conservative methods achieve significantly higher rewards and lower harm rates, demonstrating their relative superiority. 


\begin{figure}[!t]  
\centering
\subfloat[$\hat R(\hat \pi)$, $\lambda = 0.05$]{
\begin{minipage}[t]{0.4 \linewidth} 
\centering
\includegraphics[width=1\textwidth]{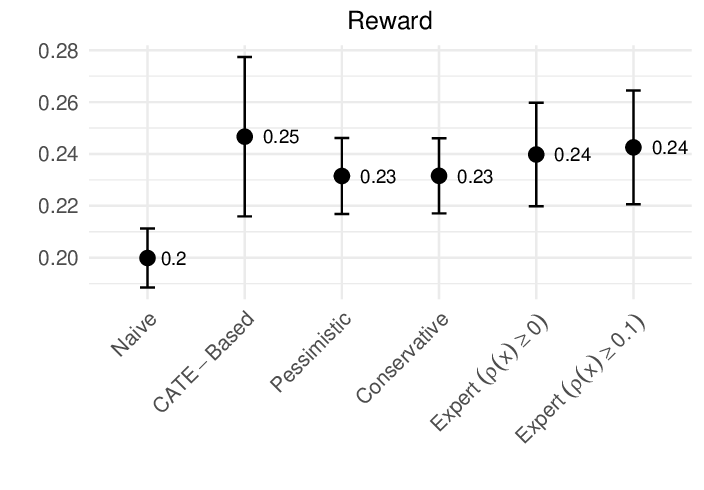}
\end{minipage}%
 \vspace{-15pt} }  \hspace{-20pt} 
\vspace{-5pt}  
\subfloat[$\hat\E(\hat \pi)$, $\lambda = 0.05$]{
\begin{minipage}[t]{0.4 \linewidth}
\centering
\includegraphics[width=1\textwidth]{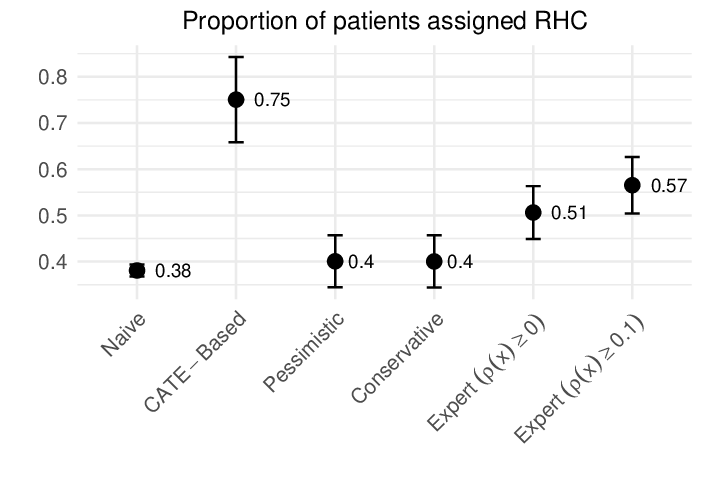}
\end{minipage} 
 \vspace{-15pt} } 

\subfloat[$\widehat{\text{THR}}_1(\hat \pi)$, $\lambda = 0.05$]{
\begin{minipage}[t]{0.32 \linewidth} 
\centering
\includegraphics[width=1\textwidth]{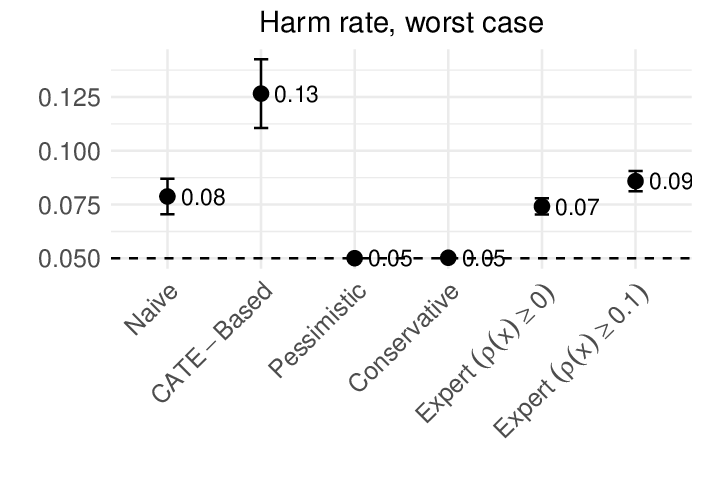}
\end{minipage}%
 \vspace{-10pt}} \hspace{-10pt} 
\subfloat[$\widehat{\text{THR}}_2(\hat \pi)$, $\lambda = 0.05$]{
\begin{minipage}[t]{0.32 \linewidth}
\centering
\includegraphics[width=1\textwidth]{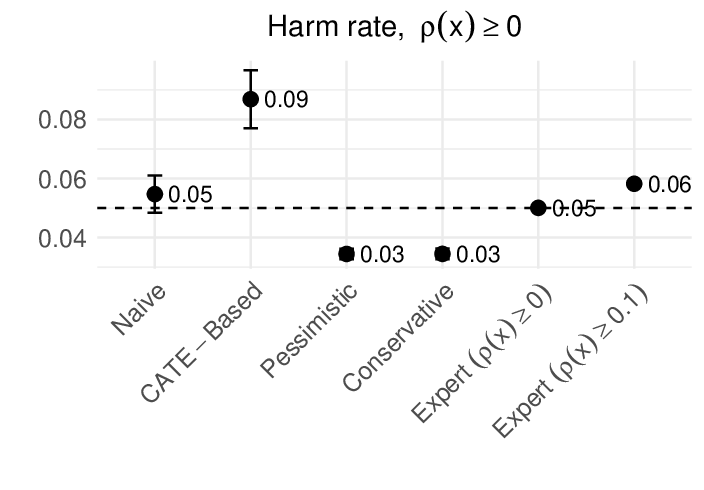}
\end{minipage} 
 \vspace{-10pt}} \hspace{-10pt} 
\subfloat[$\widehat{\text{THR}}_3(\hat \pi)$, $\lambda = 0.05$]{
\begin{minipage}[t]{0.32 \linewidth}
\centering
\includegraphics[width=1\textwidth]{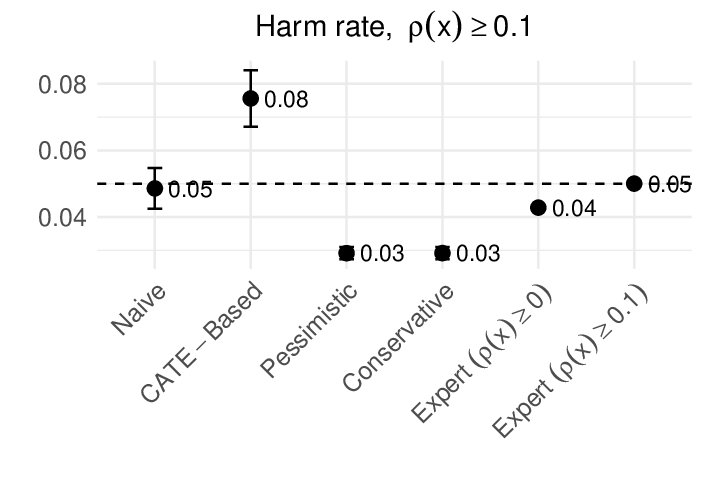}
\end{minipage}  } 
\caption{Point estimates and 95\% CIs for various evaluation metrics in the SUPPORT.} 
\label{fig2} %
\end{figure}

\begin{figure}[!h]
\centering
\subfloat[$\hat R(\hat \pi)$, $\lambda = 0.02$]{
\begin{minipage}[t]{0.4 \linewidth} 
\centering
\includegraphics[width=1\textwidth]{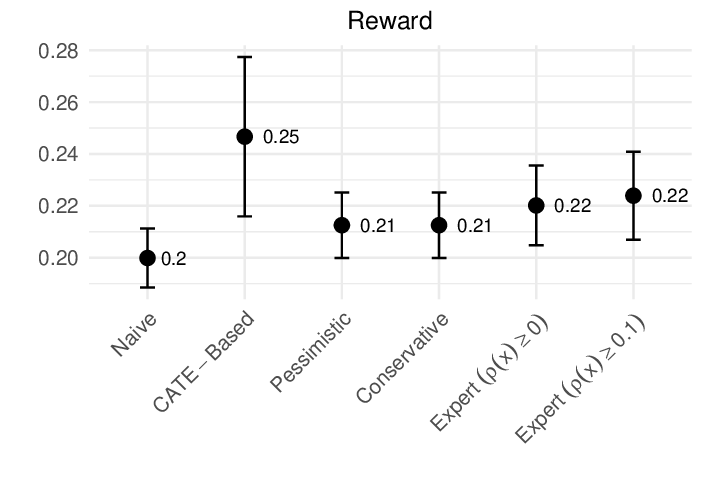}
\end{minipage}%
 \vspace{-15pt}}   \hspace{-20pt} 
\subfloat[$\hat\E(\hat \pi)$, $\lambda = 0.02$]{
\begin{minipage}[t]{0.4 \linewidth}
\centering
\includegraphics[width=1\textwidth]{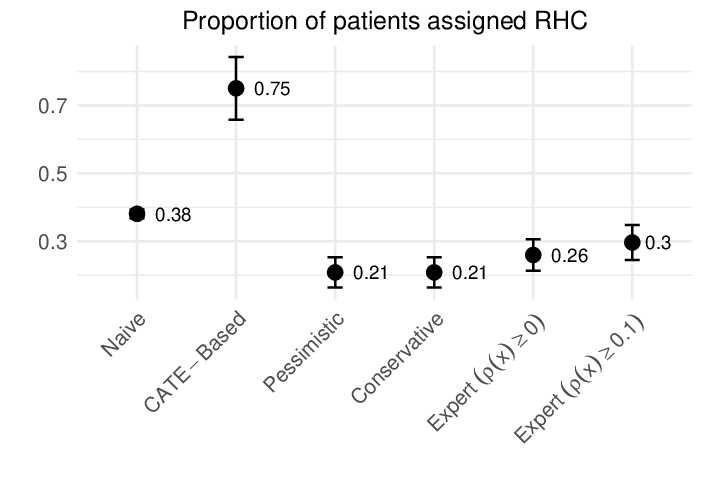}
\end{minipage} 
 \vspace{-10pt}} 

\subfloat[$\widehat{\text{THR}}_1(\hat \pi)$, $\lambda = 0.02$]{
\begin{minipage}[t]{0.32 \linewidth} 
\centering
\includegraphics[width=1\textwidth]{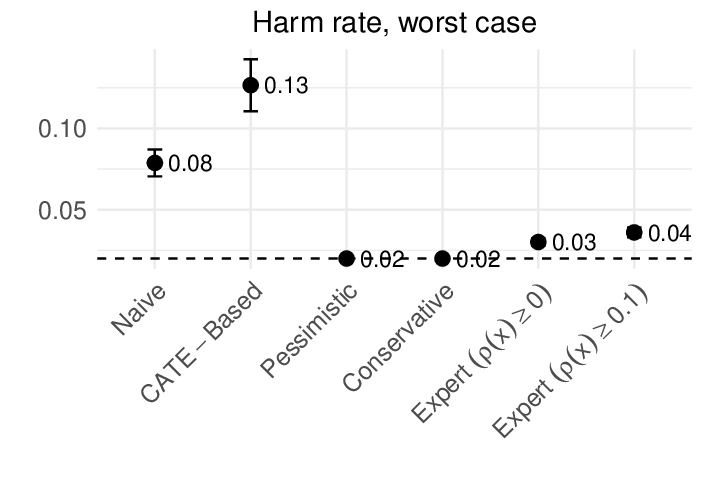}
\end{minipage}%
 \vspace{-10pt}} \hspace{-10pt} 
\subfloat[$\widehat{\text{THR}}_2(\hat \pi)$, $\lambda = 0.02$]{
\begin{minipage}[t]{0.32 \linewidth}
\centering
\includegraphics[width=1\textwidth]{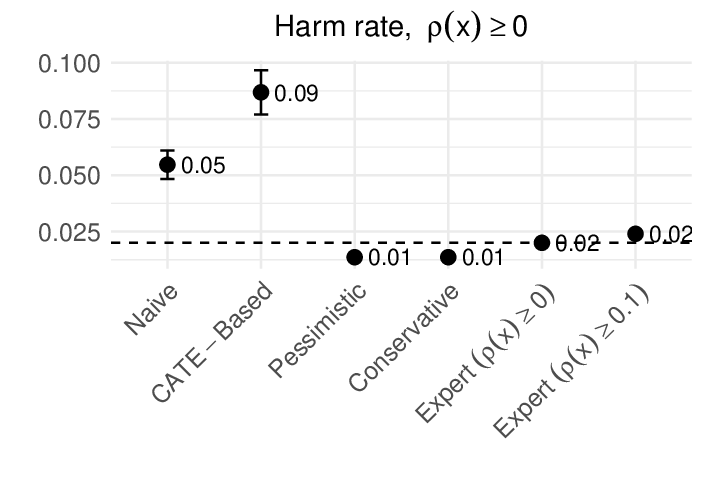}
\end{minipage} 
 \vspace{-10pt}}  \hspace{-10pt} 
\subfloat[$\widehat{\text{THR}}_3(\hat \pi)$, $\lambda = 0.02$]{
\begin{minipage}[t]{0.32 \linewidth}
\centering
\includegraphics[width=1\textwidth]{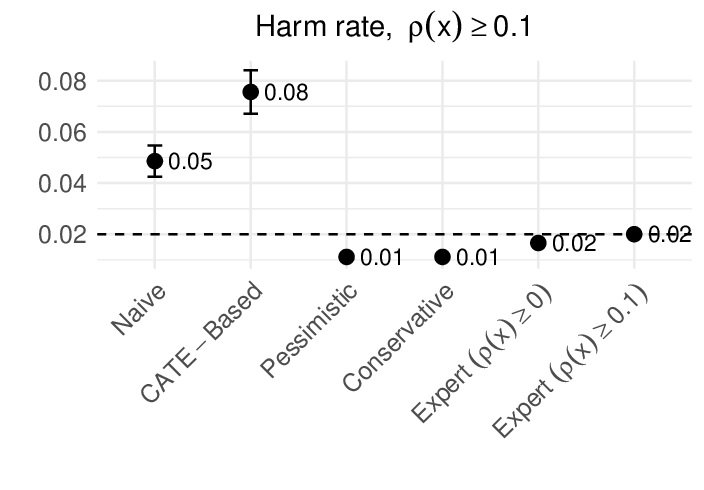}
\end{minipage} 
 \vspace{-10pt}} 
\caption{Point estimates and 95\% CIs for various evaluation metrics in the SUPPORT.} 
\label{fig3} 
\end{figure}

Figure~\ref{fig3} presents the evaluation metrics under $\lambda = 0.02$, from which it is evident that the Naive and CATE-Based methods fail to control the harm rate below the specified threshold. In contrast, the Pessimistic and Conservative methods substantially reduce the harm rate; however, this comes at the cost of assigning a significantly smaller proportion of patients to receive RHC. The  {Expert ($\rho(x)\geq 0$) and Expert ($\rho(x)\geq 0.1$)}  methods achieve a more balanced performance: they considerably reduce the harm rate while maintaining relatively high rewards and assigning a moderate proportion of patients to receive RHC. 


\subsection{Safe ITRs in the \emph{J\'{o}venes en Acci\'{o}n}}

{\bf Background.}  The primary objectives of \emph{J\'{o}venes en Acci\'{o}n} were to strengthen labor market attachment and improve job quality for disadvantaged young individuals in Colombia. 
\citet{Attanasio-etal2011} analyzed the short-run effects of the program, reporting improvements in employability and job quality. Subsequently, \citet{Attanasio-etal2017} further demonstrated that the program's effects persist in a long-run follow-up using an experimental evaluation sample.  
\emph{We use the evaluation sample to learn safe ITRs and to assess the potential reward uplift and harm reduction relative to random assignment in the dataset.} 
The treatment $A$ indicates whether an individual attended the vocational training, and $Y$ indicates whether they secured formal employment at follow-up.  
The covariates $X$ consist of 12 variables capturing demographic characteristics and economic status at baseline.  After excluding individuals lost to follow-up, the analytical dataset comprises 2,998 individuals.

\smallskip
\noindent 
{\bf Results.} We present the results in Figures~\ref{fig4} and \ref{fig5}, with \(\lambda = 0.10\) and \(\lambda = 0.05\) as the harm rate thresholds, respectivley.   
From these results, we have the following observation: the aggressive CATE-based method assigns 98\% of individuals to vocational training, indicating that the estimated CATE is positive for most of them;  The Naive method induces a harm rate ranging from 0.078 to 0.134;  When $\lambda = 0.10$, the Pessimistic and Conservative methods perform better than the Naive method in terms of higher reward, lower harm rates, and a slightly lower proportion of individuals assigned to vocational training;  When $\lambda = 0.05$, the Naive method induces a harm rate that significantly exceeds the threshold. In contrast, the Expert ($\rho(x)\geq 0$) and Expert ($\rho(x)\geq 0.1$) methods achieve more balanced performance: they substantially reduce harm rates while maintaining relatively high rewards and assigning a moderate proportion of individuals to vocational training.

\begin{figure}[!t]  
\centering
\subfloat[$\hat R(\hat \pi)$, $\lambda = 0.10$]{
\begin{minipage}[t]{0.4 \linewidth} 
\centering
\includegraphics[width=1\textwidth]{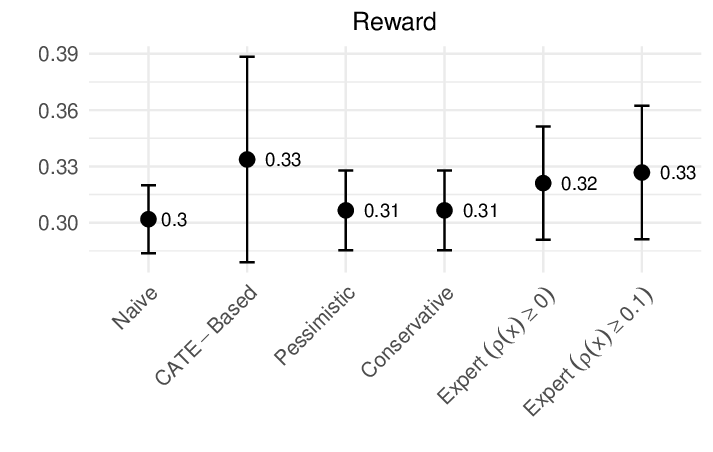}
\end{minipage}%
\vspace{-10pt}  }  \hspace{-20pt} 
\vspace{-2pt}  
\subfloat[$\hat\E(\hat \pi)$, $\lambda = 0.10$]{
\begin{minipage}[t]{0.4 \linewidth}
\centering
\includegraphics[width=1\textwidth]{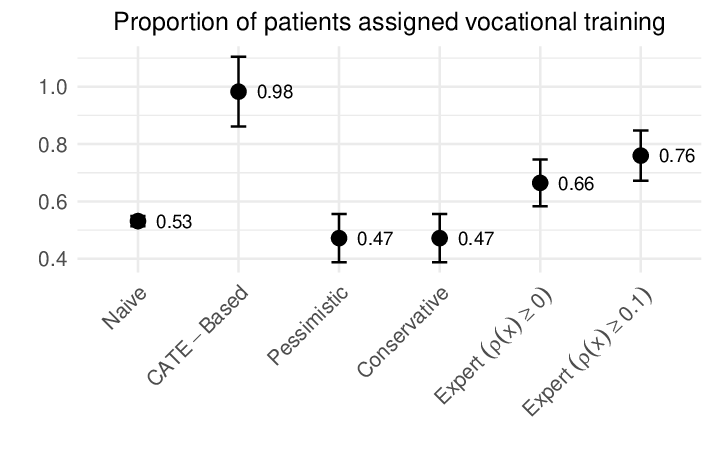}
\end{minipage} 
\vspace{-10pt} } 
\vspace{15pt}  
\subfloat[$\widehat{\text{THR}}_1(\hat \pi)$, $\lambda = 0.10$]{
\begin{minipage}[t]{0.32 \linewidth} 
\centering
\includegraphics[width=1\textwidth]{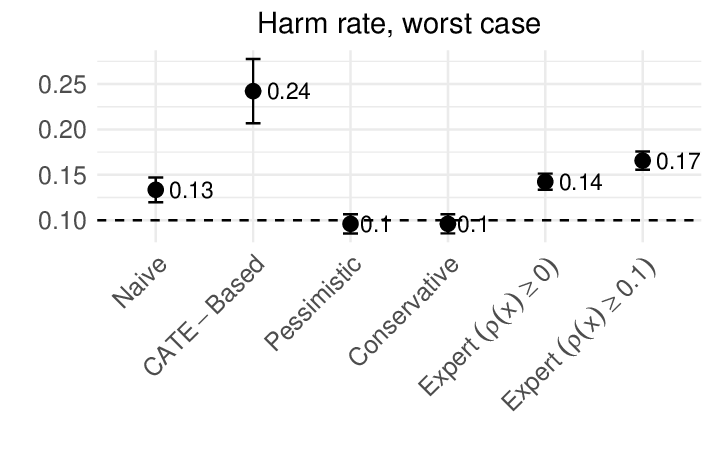}
\end{minipage}%
\vspace{-10pt} } \hspace{-10pt} 
\subfloat[$\widehat{\text{THR}}_2(\hat \pi)$, $\lambda = 0.10$]{
\begin{minipage}[t]{0.32 \linewidth}
\centering
\includegraphics[width=1\textwidth]{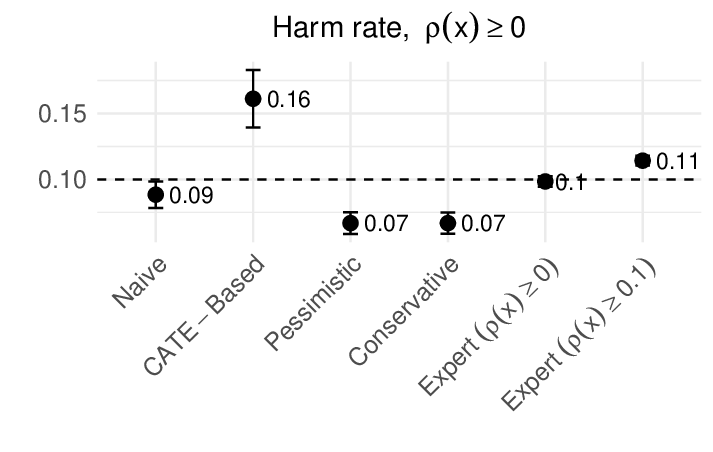}
\end{minipage} 
\vspace{-10pt} }  \hspace{-10pt} 
\subfloat[$\widehat{\text{THR}}_3(\hat \pi)$, $\lambda = 0.10$]{
\begin{minipage}[t]{0.32 \linewidth}
\centering
\includegraphics[width=1\textwidth]{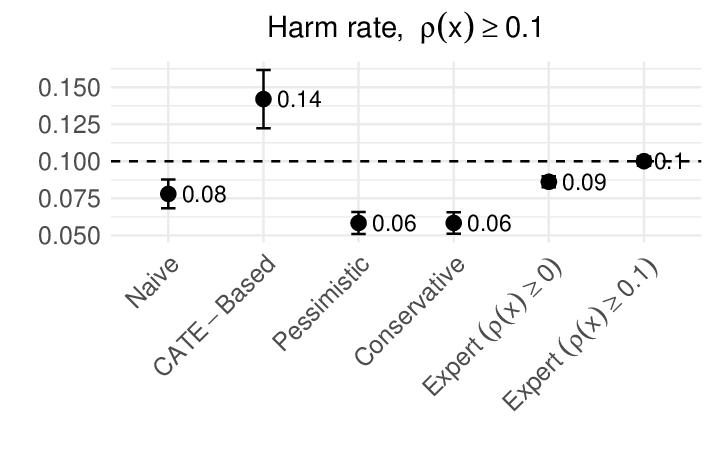}
\end{minipage} 
\vspace{-10pt} } 
\caption{Point estimates and 95\% CIs in the \emph{J\'{o}venes en Acci\'{o}n}.} 
\label{fig4} 
\end{figure}

\begin{figure}[!h]
\centering
\subfloat[$\hat R(\hat \pi)$, $\lambda = 0.05$]{
\begin{minipage}[t]{0.4 \linewidth} 
\centering
\includegraphics[width=1\textwidth]{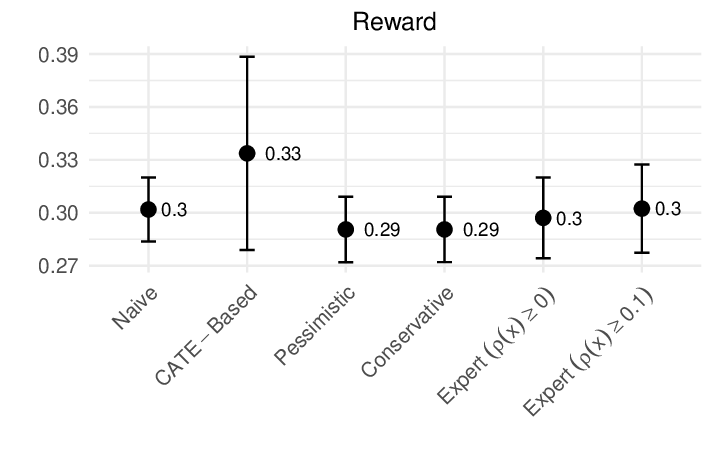}
\end{minipage}%
\vspace{-15pt} }    \hspace{-20pt} 
\vspace{-5pt}  
\subfloat[$\hat\E(\hat \pi)$, $\lambda = 0.05$]{
\begin{minipage}[t]{0.4 \linewidth}
\centering
\includegraphics[width=1\textwidth]{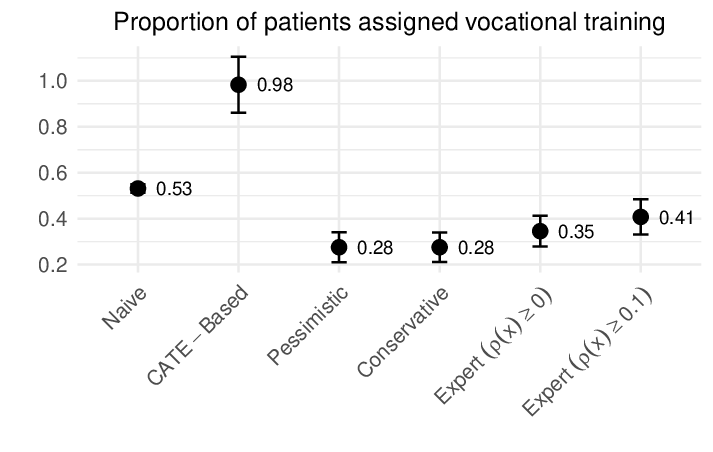}
\end{minipage} 
\vspace{-15pt} 
} 
\vspace{15pt}  
\vspace{-5pt}  
\subfloat[$\widehat{\text{THR}}_1(\hat \pi)$, $\lambda = 0.05$]{
\begin{minipage}[t]{0.32 \linewidth} 
\centering
\includegraphics[width=1\textwidth]{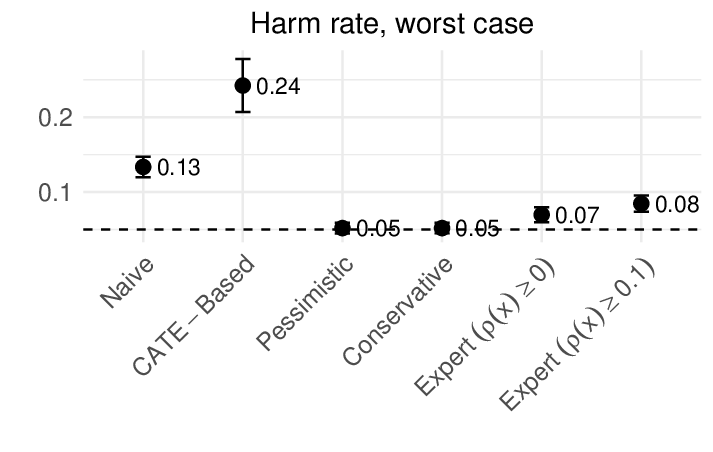}
\end{minipage}%
\vspace{-10pt} } \hspace{-10pt}  
\subfloat[$\widehat{\text{THR}}_2(\hat \pi)$, $\lambda = 0.05$]{
\begin{minipage}[t]{0.32 \linewidth}
\centering
\includegraphics[width=1\textwidth]{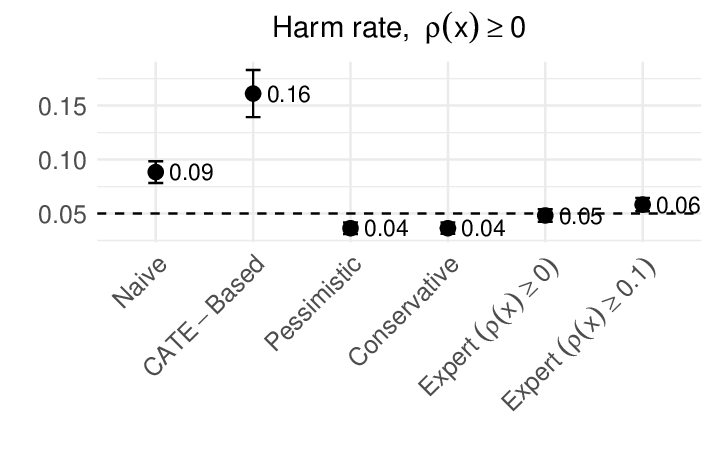}
\end{minipage} 
\vspace{-10pt} }  \hspace{-10pt} 
\subfloat[$\widehat{\text{THR}}_3(\hat \pi)$, $\lambda = 0.05$]{
\begin{minipage}[t]{0.32 \linewidth}
\centering
\includegraphics[width=1\textwidth]{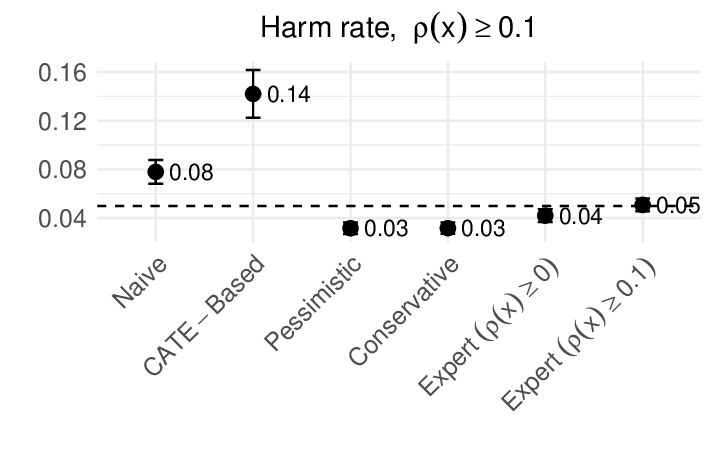}
\end{minipage} } 
\hspace{-10pt}
\caption{Point estimates and 95\% CIs in the \emph{J\'{o}venes en Acci\'{o}n}.} 
\label{fig5} %
\end{figure}


\section{Conclusion}
\label{sec:conclusion}

In this article, we propose a suite of methods for estimating ITRs that aim to maximize reward while ensuring that the induced harm rate remains below a pre-specified threshold. We derived the explicit form of the theoretically optimal ITR and developed several estimation strategies for ITRs under partial identification of harm rates. By jointly accounting for reward maximization and harm control, our proposed methods offer practical and user-friendly solutions for estimating ITRs, particularly well-suited to high-stakes decision-making scenarios where mitigating potential harm is of paramount importance. 


%
The proposed methods focus on binary outcomes. A natural extension is to consider ordinal or continuous outcomes. In such settings, for a given ITR $\pi$, it is necessary to account not only for the induced harm rate,  
$\E\{\P(Y(1) - Y(0) < 0 \mid X=x) \pi(X)\},$  
but also for the harm quantity, $\E\left( \E[\{Y(0)-Y(1)\} \mathbb{I}\{Y(1) - Y(0) < 0\} \mid X ] \pi(X)\right).$  
Moreover, in certain highly risk-averse scenarios, it may be of interest to minimize the harm rate as much as possible while ensuring that the benefit rate exceeds a predefined threshold. This objective can be formulated as the problem of estimating ITRs that solve $\min_{\pi} \text{THR}(\pi) \ \text{s.t.} \ R(\pi) \geq \lambda$. A discussion of this optimization problem is provided in the Supplementary Material.    



\section*{Acknowledgement}
We thank Prof. Qingyuan Zhao and Dr. Yue Zhang for their valuable discussions.


\bibliographystyle{plainnat}
\bibliography{paper-ref}
 


%

\newpage

  \begin{center}
\bf \Large 
Supplementary Material
\end{center}

\setcounter{equation}{0}
\setcounter{section}{0}
\setcounter{figure}{0}
\setcounter{example}{0}
\setcounter{proposition}{0}
\setcounter{corollary}{0}
\setcounter{theorem}{0}
\setcounter{table}{0}
\setcounter{condition}{0}
\setcounter{lemma}{0}
\setcounter{remark}{0}

\renewcommand {\theproposition} {S\arabic{proposition}}
\renewcommand {\theexample} {S\arabic{example}}
\renewcommand {\thefigure} {S\arabic{figure}}
\renewcommand {\thetable} {S\arabic{table}}
\renewcommand {\theequation} {S\arabic{equation}}
\renewcommand {\thelemma} {S\arabic{lemma}}
\renewcommand {\thesection} {S\arabic{section}}
\renewcommand {\thetheorem} {S\arabic{theorem}}
\renewcommand {\thecorollary} {S\arabic{corollary}}
\renewcommand {\thecondition} {S\arabic{condition}}
\renewcommand {\thepage} {S\arabic{page}}
\renewcommand {\theremark} {S\arabic{remark}}

\setcounter{page}{1}

  \setcounter{equation}{0}
\renewcommand {\theequation} {S\arabic{equation}}
  \setcounter{lemma}{0}
\renewcommand {\thelemma} {S\arabic{lemma}}
   \setcounter{definition}{0}
\renewcommand {\thedefinition} {S\arabic{definition}}
   \setcounter{example}{0}
\renewcommand {\theexample} {S\arabic{example}}
   \setcounter{proposition}{0}
\renewcommand {\theproposition} {S\arabic{proposition}}
   \setcounter{corollary}{0}
\renewcommand {\thecorollary} {S\arabic{corollary}}

 \bigskip 
 
The Supplementary Material contains proofs of the theoretical results, an additional estimation method, and additional simulation studies in the main paper.

\section{Technical Proofs}

\subsection{Proofs of Theorem \ref{thm1}}

\emph{Proof of Theorem \ref{thm1}}. 
First, we note that 
   \begin{align*}
     \Pi^*_\lambda ={}& \begin{cases}
\arg \max_{\pi} \quad \E[\pi(X) Y(1) + (1 - \pi(X))Y(0)]\\
s.t. \quad  \text{THR}(\pi) \leq \lambda\\
\end{cases}   \\
   ={}&  \begin{cases} 
          \arg \max_{\pi} \quad \E[\pi(X) (Y(1) - Y(0))] + \E[Y(0)]\\
s.t. \quad   \E[\text{THR}(X) \cdot \pi(X) ] \leq \lambda\\
        \end{cases} \\
           ={}&  \begin{cases} 
          \arg \max_{\pi} \quad \E[\pi(X) \tau(X) ]  \\
s.t. \quad   \E[\text{THR}(X) \cdot \pi(X) ] \leq \lambda\\
        \end{cases} 
   \end{align*}

On the one hand, if $\lambda \geq \E[\text{THR}(X) \cdot \mathbb{I}(\tau(X) > 0)]$, then the threshold $\lambda$  is higher than the maximum harm rate induced by the optimal ITR $\mathbb{I}(\tau(X) > 0)$ without the constraint (see Remark \ref{Remark1} of the manuscript for more details). In this case, the constraint $ \E[\text{THR}(X) \cdot \pi(X) ] \leq \lambda$ in the optimization problem \eqref{eq1} is automatically satisfied by maximizing only the reward, and the constrained optimization problem turns into an unconstrained optimization problem. This leads to Theorem \ref{thm1}(b). 

   \begin{table}[h!] 
\centering
  \centering
  \begin{tabular}{cccc}
    \toprule
                 &  $\tau(X) > 0$ &  $\tau(X) \leq 0 $ \\
    \midrule 
     $\text{THR}(X) > 0$  & $\mathcal{G}_1$ & $\mathcal{G}_2$ \\
    $\text{THR}(X) = 0 $  & $\mathcal{G}_3$ & $\mathcal{G}_4$ \\
    \bottomrule
  \end{tabular}
\end{table}

On the other hand,  if $\lambda < \E[ \text{THR}(X)  \cdot \mathbb{I}(\tau(X) > 0)]$, let $\pi^*_\lambda$ be an ITR in $\Pi_\lambda^*$, 
 we show that the optimal ITR $\pi^*_\lambda$ must satisfy $\E[\text{THR}(X) \cdot \pi^*_\lambda(X) ] = \lambda$, i.e., the optimal solution is obtained at the boundary of the constraint. We prove this using the method of contradiction. If we assume that the optimal ITR is not obtained at the boundary of the constraint, i.e., 
$\E[\text{THR}(X) \cdot \pi^*_\lambda(X) ] < \lambda$, 
  then when  $\lambda < \E[ \text{THR}(X)  \cdot \mathbb{I}(\tau(X) > 0)]$,    
  there are some individuals in the group $\mathcal{G}_1$ who are not assigned to treatment under $\pi^*_\lambda$. Thus, we could find another ITR $\tilde \pi^*_\lambda$
  that assigns treatment to some of the individuals in  $\mathcal{G}_1$ who were not assigned by $\pi^*_\lambda$. Compared with the optimal ITR $\pi_{\lambda}^*$,  $\tilde \pi^*_\lambda$ yields a higher reward but increases the harm rate. That is, the ITR $\tilde \pi^*_\lambda$ will lead to a harm rate closer to $\lambda$  but will have a higher reward than  $\pi^*_\lambda$; thus, $\pi^*_\lambda$ is not the optimal ITR, which contradicts its definition of $\pi^*_\lambda$.  
Therefore,  
   \begin{align*}
     \Pi^*_\lambda  
           ={}&  \begin{cases} 
          \arg \max_{\pi} \quad \E[\pi(X) \tau(X) ]  \\
s.t. \quad   \E[\text{THR}(X) \cdot \pi(X) ] =\lambda. 
        \end{cases} 
   \end{align*} 
By introducing the Lagrange multiplier $\beta$, $\pi^*_\lambda$ satisfies 
\[ \begin{cases}
\arg \max_{\pi} \quad \E[\pi(X) (\tau(X) - \beta \text{THR}(X))] \\
s.t. \quad   \E[\text{THR}(X)\cdot \pi(X)] = \lambda. 
\end{cases} \]  
This implies Theorem \ref{thm1}(a).  

\hfill $\Box$


\subsection{Proof of Lemma \ref{lem1}}

The following Lemmas \ref{lem-s1} and \ref{lem-s2} will be used in the proofs of Lemma \ref{lem1}, Theorem \ref{thm2}, and Theorem \ref{thm3}. 

\begin{lemma} \label{lem-s1} Let $\hat f$ and $f$ take any real values. Then
		\begin{align*}    & \big |  \mathbb{I}(\hat  f > 0) -  \mathbb{I}(f > 0 ) \big | \leq  \mathbb{I}( |f| \leq |\hat f - f| ). 
		\end{align*}
\end{lemma}
\emph{Proof of Lemma \ref{lem-s1}}.   Note that 
	\[  \big |  \mathbb{I}(\hat  f > 0) -  \mathbb{I}(f > 0 ) \big | = \mathbb{I}(\hat f, f \text{ have opposite sign})     \]
and if $\hat f$ and $f$ have the opposite sign then 
	\[   |\hat f|  + |f|  = |\hat f - f |,     \]
which implies that $|f| \leq |\hat f - f| $. Thus, whenever $   \big |  \mathbb{I}(\hat  f > 0) -  \mathbb{I}(f > 0 ) \big | = 1$, it must have $\mathbb{I}( |f| \leq |\hat f - f| ) = 1$. This finishes the proof. 

\hfill $\Box$

\begin{lemma} \label{lem-s2} 
Under Condition \ref{cond2}, there exists a constant $C$ such that for all $t > 0$,
	\[  \P\big (  | \tau(X) - \beta \cdot \textup{THR}_U(X) |  \leq t  \big ) \leq C  t. \]    
\end{lemma} 
\emph{Proof of Lemma \ref{lem-s2}}. Let $g(x)$ be the density function of the random variable $\tau(X) - \beta \textup{THR}_U(X)$. Then $g(x)$ is bounded under Condition  \ref{cond2}. Without loss of generality, assume that $|g(x)| \leq C/2$ for a constant $C$. Thus, we have 
   \begin{align*}
             &     \P\big (   | \tau(X) - \beta \cdot \textup{THR}_U(X) |  \leq t  \big ) \\
             ={}& \P\big (  -t \leq   \tau(X) - \beta \cdot \textup{THR}_U(X)  \leq t  \big ) \\
          ={}& \int_{-t}^t  g(x) dx    \cdot \frac{C}{2} \cdot 2 t  = C t. 
   \end{align*}
This completes the proof. 

\hfill $\Box$

\bigskip 
Condition \ref{cond2} in the manuscript is a specific case of the margin condition below. 

\begin{condition}[Margin condition] \label{cond-s1}
The random variable $\tau(X) - \beta \textup{THR}_U(X)$ has absolutely continuous cumulative distribution function and there exist constants $C_0$ and $\alpha > 0$ such that  for all $t > 0$, 
	\[ \P\big (  | \tau(X) - \beta \cdot \textup{THR}_U(X) |  \leq t  \big ) \leq C_0  t^{\alpha}. \]    
\end{condition} 

In Condition \ref{cond-s1}, the case where $\alpha = 0$ is trivial (no assumption) and included for notational convenience.  If the density of $\tau(X) - \beta \text{THR}_U(X)$ is bounded, then by Lemma \ref{lem-s2}, Condition \ref{cond-s1} holds for $\alpha = 1$, which is precisely the case of Condition \ref{cond2} in the manuscript.  
Essentially, the parameter $\alpha$ in Condition \ref{cond-s1} determines how many cases are allowed to be close to the boundary, with a larger value meaning fewer cases are close~\citep{2018Who, Eli-etal2023}. Condition \ref{cond-s1} provides a useful characterization of the behavior of $\tau(X) - \beta \text{THR}_U(X)$ in the vicinity of the level $\tau(X) - \beta \text{THR}_U(X) = 0$, which is crucial for addressing the discontinuity of the estimator of $\beta^\dag$. The margin condition has been extensively studied and has proven useful in classification problems~\citep{Audibert-etal-2007, Laan-Luedtke2014} and other scenarios involving the estimation of non-smooth functions~\citep{Luedtke-etal-2016, Kennedy-2019}.

  \bigskip
Next, we provide a detailed proof of Lemma \ref{lem1}. Thereafter, without loss of generality, in Algorithm \ref{algorithm1} of the manuscript, we let the index set $\mathcal{I}_k$ be $\{1, 2, \ldots, n\}$, where $n = N/K$ and $K$ is a fixed constant. For ease of presentation, we denote $\hat{\beta}^{(k)}$ by $\hat{\beta}$, $\hat \tau^{(-k)}(x)$ by $\hat \tau(x)$, and $\widehat{\text{THR}}^{(-k)}_U(x)$ by $\widehat{\text{THR}}_U(x)$. Then, we write $\hat \pi_\lambda^\dag(X)$ as $ \I(\hat \tau(X) - \hat \beta \widehat{\text{THR}}_U(X) >0 )$. 
In addition, $a_n \lesssim b_n$ means that $a_n \leq b_n$ for some constant $C_1 > 0$. 

\bigskip  \noindent 
\emph{Proof of Lemma \ref{lem1}}. 
We denote $\bm{\eta}(x) := (\tau(x), \text{THR}_U(x))$ as the nuisance parameters, $\hat{\bm{\eta}}(x) := ( \hat \tau(x), \widehat{\text{THR}}_U(x))$ as the corresponding estimators, and denote
  \begin{align*}
   \Psi_n(\beta; \hat{\bm{\eta}}) :={}& n^{-1} \sum_{i=1}^n \widehat{\text{THR}}_U(X_i) \cdot  \I(\hat \tau(X_i) - \beta \cdot \widehat{\text{THR}}_U(X_i) > 0) - \lambda \\
     \Psi(\beta; \bm{\eta}) :={}&    \E[ \text{THR}_U(X) \cdot  \I(\tau(X) - \beta \cdot \text{THR}_U(X) > 0)] - \lambda.     
      \end{align*}
 Then,  $\hat \beta$ is the solution of the estimating equation $\Psi_n(\beta; \hat{\bm{\eta}}) = 0$, i.e., 
      $\Psi_n(\hat \beta; \hat{\bm{\eta}}) = 0$. Likewise, the true value $\beta^\dag$ satisfies 
      $\Psi(\beta^\dag; \bm{\eta}) = 0$.

Following similar arguments to those in the proof of Theorem 5.9 in \citet{vdv-1998}, to show the consistency of $\hat \beta$,  it suffices to verify the following two conditions: 
  \[  \sup_{\beta} | \Psi_n(\beta; \hat{\bm{\eta}}) - \Psi(\beta;  \bm{\eta})  |  \xrightarrow{\P} 0,     \] 
and for every $\epsilon > 0$ 
  \[    \inf_{\beta: | \beta - \beta^\dag | > \epsilon} | \Psi(\beta;  \bm{\eta}) | > 0 = | \Psi(\beta^\dag;  \bm{\eta)} |.   \] 
The second condition holds naturally as $\Psi(\beta;  \bm{\eta})$ is a monotone function of $\beta$ and has a unique solution at $\beta^\dag$ under Condition \ref{cond1}.

Next, we focus on verifying the first condition. Note that 
\[  \sup_{\beta} | \Psi_n(\beta; \hat{\bm{\eta}}) - \Psi(\beta; \bm{\eta})  | \leq  \sup_{\beta} | \Psi_n(\beta;  \hat{\bm{\eta}}) - \Psi_n(\beta; \bm{\eta}) |  +  \sup_{\beta} | \Psi_n(\beta;  \bm{\eta}) - \Psi(\beta; \bm{\eta}) |.      \]
By Theorem 2.7.5 of \citet{vdv-1996}, the class of bounded monotone functions has a finite bracketing integral and is thus a $\P$-Glivenko-Cantelli class, which leads to that 
\[ \sup_{\beta} | \Psi_n(\beta;  \bm{\eta}) - \Psi(\beta; \bm{\eta}) | = o_{\P}(1). \]
We then analyze $\sup_{\beta} | \Psi_n(\beta;  \hat{\bm{\eta}}) - \Psi_n(\beta; \bm{\eta}) |$. 
We decompose $\Psi_n(\beta;  \hat{\bm{\eta}}) - \Psi_n(\beta; \bm{\eta}) $ as follows 
\[  \Psi_n(\beta;  \hat{\bm{\eta}}) - \Psi_n(\beta; \bm{\eta})  =  A_{1n}(\beta) + A_{2n}(\beta),   \] 
where 
\begin{align*}
A_{1n}(\beta) ={}&   \frac 1 n  \sum_{i=1}^n  \widehat{\text{THR}}_U(X_i) \cdot \I(  \hat \tau(X_i) - \beta \cdot \widehat{\text{THR}}_U(X_i) > 0 ) \\
{}&  -   \frac 1 n  \sum_{i=1}^n  \text{THR}_U(X_i) \cdot \I(  \hat \tau(X_i) - \beta \cdot \widehat{\text{THR}}_U(X_i) > 0 )  \\
A_{2n}(\beta) ={}&   \frac 1 n  \sum_{i=1}^n \text{THR}_U(X_i) \cdot \I(  \hat \tau(X_i) - \beta \cdot \widehat{\text{THR}}_U(X_i) > 0 ) \\
{}& -   \frac 1 n  \sum_{i=1}^n   \text{THR}_U(X_i) \cdot \I(   \tau(X_i) - \beta \cdot \text{THR}_U(X_i) > 0 ) \Big \}. 
\end{align*}
For $A_{1n}(\beta)$, we have that
\[  \sup_{\beta} |A_{1n}(\beta)| \leq   \frac 1 n  \sum_{i=1}^n \Big   | \widehat{\text{THR}}_U(X_i) -   \text{THR}_U(X_i) \Big |  \leq \sup_x |  \widehat{\text{THR}}_U(x) - \text{THR}_U(x) |  = o_{\P}(1).  \] 
For $A_{2n}(\beta)$, we have that 
\begin{align*}
& \sup_{\beta} |A_{2n}(\beta)|  \\
\leq{}&   \frac 1 n  \sum_{i=1}^n  \text{THR}_U(X_i) \cdot \Big | \I(  \hat \tau(X_i) - \beta \cdot \widehat{\text{THR}}_U(X_i) > 0 ) -  \I(   \tau(X_i) - \beta \cdot \text{THR}_U(X_i) > 0 )  \Big |  \\
\leq{}& \frac 1 n  \sum_{i=1}^n \Big | \I(  \hat \tau(X_i) - \beta \cdot \widehat{\text{THR}}_U(X_i) > 0 ) -  \I(   \tau(X_i) - \beta \cdot \text{THR}_U(X_i) > 0 )  \Big | \\
\leq{}&  \frac 1 n  \sum_{i=1}^n  \I\Big ( \big |   \tau(X_i) - \beta \cdot \text{THR}_U(X_i) \big | \leq \big |    \hat \tau(X_i) - \tau(X_i) - \beta \cdot (\widehat{\text{THR}}_U(X_i) - \text{THR}_U(X_i) ) \big |  \Big ) \\
={}&  \P\Big ( \big |   \tau(X_i) - \beta \cdot \text{THR}_U(X_i) \big | \leq \big |    \hat \tau(X_i) - \tau(X_i) - \beta \cdot (\widehat{\text{THR}}_U(X_i) - \text{THR}_U(X_i) ) \big |   \Big ) + o_{\P}(1)    \\
\lesssim{}&   \sup_{\beta} \sup_{x} \big |  \hat \tau(x) - \tau(x) - \beta \cdot (\widehat{\text{THR}}_U(x) - \text{THR}_U(x) ) \big | + o_{\P}(1) \\
={}& o_{\P}(1),
\end{align*}
where the second inequality follows from $0\leq \text{THR}_U(X)\leq 1$, the third inequality follows from Lemma \ref{lem-s1},  the fourth inequality follows from  Lemma \ref{lem-s2}.   
Therefore, $\sup_{\beta} |\Psi_n(\beta;  \hat{\bm{\eta}}) - \Psi_n(\beta; \bm{\eta}) | = o_{\P}(1)$, and the first condition is verified. This completes the proof.  
  
  \hfill $\Box$


\subsection{Proof of Theorem \ref{thm2}}

\emph{Proof of Theorem \ref{thm2}}. We prove Theorems \ref{thm2}(a) and \ref{thm2}(b), respectively.

Theorem \ref{thm2}(a) holds immediately by noting that 
\begin{align*}
&  \E[ |\hat \pi_\lambda^\dag(X) - \pi^\dag_\lambda(X) | ] \\
={}& \E \left [ \big | \I(\hat \tau(X) - \hat \beta \widehat{\text{THR}}(X) >0 ) - \I( \tau(X) - \beta^\dag \text{THR}_U(X) >0 ) \big  | \right ] \\
\leq{}&  \E \left [  \I \Big ( |\tau(X) - \beta^\dag \text{THR}_U(X)| \leq | \{\hat \tau(X) - \tau(X)\} - \{\hat \beta \widehat{\text{THR}}(X) - \beta^\dag \text{THR}_U(X) \} | \Big )     \right ] \\
\lesssim{}&    \{ \sup_x |\hat \tau(x) - \tau(x)|\} +  \{ \sup_x |\hat \beta \widehat{\text{THR}}_U(x) - \beta^\dag \text{THR}_U(x)| \}  \\
={}& o_{\P}(1),
\end{align*}
where the first inequality follows from Lemma \ref{lem-s1}, the second inequality from Lemma \ref{lem-s2}, and the last  equality from Lemma \ref{lem1} and the uniform consistency of $\hat \tau(x)$ and $\widehat{\textup{THR}}_U(x)$. 

Next, we prove Theorem \ref{thm2}(b). Similar to the proof of Theorem \ref{thm2}(a), we have that 
\begin{align*}
& U(\hat \pi_\lambda) - U(\pi^\dag_\lambda) \\
={}&   \E[ (\hat \pi_\lambda^\dag(X) - \pi^\dag_\lambda(X) ) \cdot (\tau(X) - \beta^\dag {\text{THR}}_U(X))] \\
\leq{}&  \E \left [ \big | \hat \pi_\lambda^\dag(X) - \pi^\dag_\lambda(X)  \big | \cdot \big | \tau(X) - \beta^\dag {\text{THR}}_U(X) \big | \right ] \\
={}&   \E \left [ \big | \I(\hat \tau(X) - \hat \beta \widehat{\text{THR}}_U(X) >0 ) - \I( \tau(X) - \beta^\dag \text{THR}_U(X) >0 ) \big | \cdot \big | \tau(X) - \beta^\dag {\text{THR}}_U(X) \big | \right ] \\
\leq{}&   \E \Big [  \I \Big ( |\tau(X) -  \beta^\dag {\text{THR}}_U(X)|    \leq | \{\hat \tau(X) - \tau(X)\} - \{\hat \beta \widehat{\text{THR}}_U(X) - \beta^\dag \text{THR}_U(X) \} | \Big ) \\
                     {}& \times  \big | \tau(X) -  \beta^\dag {\text{THR}}_U(X)) \big | \Big ] \\
\lesssim{}&  \E[ | \{\hat \tau(X) - \tau(X)\} - \{\hat \beta \widehat{\text{THR}}_U(X) - \beta^\dag \text{THR}_U(X) \} | ] \\
={}& o_{\P}(1), 
\end{align*}  
where the last inequality (``$\lesssim$") follows from the boundedness of  $\big | \tau(X) -  \beta^\dag {\text{THR}}_U(X)) \big |$. 
This leads to the conclusion of Theorem \ref{thm2}(b). 

\hfill $\Box$

  \subsection{Proof of Theorem \ref{thm3}}

The following Lemmas \ref{lem-s3}--\ref{lem-s6} will be used to prove Theorem \ref{thm3}.

\begin{lemma} \label{lem-s3} For any plug-in ITR $\hat \pi_\lambda^\dag(X)$ with this form of $\I( \hat \tau(X) -  \hat \beta \widehat{\textup{THR}}_U(X) > 0)$, 
\[     U(\pi^\dag_\lambda) - U(\hat \pi_\lambda) = \E\Big [   | \tau(X) - \beta^\dag {\textup{THR}}_U(X) | \cdot \I( \hat \pi_\lambda^\dag(X)  \neq \pi_\lambda^\dag(X)) \Big ]        \] 
\end{lemma} 

  \noindent 
\emph{Proof of Lemma \ref{lem-s3}}.  
Note that 
\begin{align*}
&  U(\pi^\dag_\lambda) - U(\hat \pi_\lambda)  \\
={}&   \E[ (\pi^\dag_\lambda(X) - \hat \pi_\lambda^\dag(X)) \cdot (\tau(X) - \beta^\dag {\text{THR}}_U(X))] \\
={}&   \E[ (\pi^\dag_\lambda(X) - \hat \pi_\lambda^\dag(X)) \cdot (\tau(X) - \beta^\dag {\text{THR}}_U(X)) \cdot   \I( \hat \pi_\lambda^\dag(X)  \neq \pi_\lambda^\dag(X)) ]. 
\end{align*}
To show Lemma \ref{lem-s3}, it suffices to verify that 
\begin{align*}
&  | \tau(X) - \beta^\dag {\textup{THR}}_U(X) | \cdot    \I( \hat \pi_\lambda^\dag(X)  \neq \pi^\dag_\lambda(X))  \\
={}& (\pi^\dag_\lambda(X) - \hat \pi_\lambda^\dag(X)) \cdot (\tau(X) - \beta^\dag {\text{THR}}_U(X)) \cdot   \I( \hat \pi_\lambda^\dag(X)  \neq \pi^\dag_\lambda(X)). 
\end{align*}

On one hand, when $\hat \pi_\lambda^\dag(X)  = \pi^\dag_\lambda(X)$, 
\begin{align*} 
& | \tau(X) - \beta^\dag {\textup{THR}}_U(X) | \cdot    \I( \hat \pi_\lambda^\dag(X)  \neq \pi^\dag_\lambda(X)) \equiv 0, \\
& (\pi^\dag_\lambda(X) - \hat \pi_\lambda^\dag(X)) \cdot (\tau(X) - \beta^\dag {\textup{THR}}_U(X)) \cdot   \I( \hat \pi_\lambda^\dag(X)  \neq \pi^\dag_\lambda(X))  \equiv 0. 
\end{align*} 

On the other hand,  when  $\hat \pi_\lambda^\dag(X)  \neq \pi^\dag_\lambda(X)$, it is sufficient to show that 
\[  (\pi^\dag_\lambda(X) - \hat \pi_\lambda^\dag(X))  \cdot (\tau(X) - \beta^\dag {\textup{THR}}_U(X)) =  | \tau(X) - \beta^\dag {\textup{THR}}_U(X) |.  \] 
which follows immediately from the following observations: 
  \begin{itemize}
\item if  $\tau(X) - \beta^\dag {\textup{THR}}_U(X) > 0$, then $\pi^\dag_\lambda(X) = 1$ by its definition, which implies that $\hat \pi_\lambda^\dag(X) = 0$ due to   $\hat \pi_\lambda^\dag(X)  \neq \pi^\dag_\lambda(X)$. In this case, 
\[ (\pi^\dag_\lambda(X) - \hat \pi_\lambda^\dag(X)  ) \cdot (\tau(X) - \beta^\dag {\textup{THR}}_U(X))  =  \tau(X) - \beta^\dag {\textup{THR}}_U(X) = | (\tau(X) - \beta^\dag {\textup{THR}}_U(X)) |.   \]

\item if  $\tau(X) - \beta^\dag {\textup{THR}}_U(X) \leq 0$, then $\pi^\dag_\lambda(X) = 0$ and $\hat \pi_\lambda^\dag(X) = 1$. In this case, 
\[  (\pi^\dag_\lambda(X) - \hat \pi_\lambda^\dag(X)  ) \cdot (\tau(X) - \beta^\dag {\textup{THR}}_U(X))  = - \{ \tau(X) - \beta^\dag {\textup{THR}}_U(X)\} = | (\tau(X) - \beta^\dag {\textup{THR}}_U(X)) |.   \]
\end{itemize} 
This completes the proof. 

\hfill $\Box$

  \begin{lemma}  \label{lem-s4}  Under Condition \ref{cond2}, then for any plug-in ITR $\hat \pi_\lambda^\dag(X)$ with this form of $\I( \hat \tau(X) -  \hat \beta \widehat{\textup{THR}}_U(X) > 0)$, we have that 
\[    U(\pi^\dag_\lambda) - U(\hat \pi_\lambda^\dag)  \leq  C ||  \{\hat \tau(X) - \tau(X)\} - \{\hat \beta \widehat{\textup{THR}}_U(X) - \beta^\dag \textup{THR}_U(X) \}  ||_{\infty}^{2},    \]
where $C > 0$ is a constant defined in Lemma \ref{lem-s2}.    
\end{lemma}

  \noindent 
\emph{Proof of Lemma \ref{lem-s4}}. 
We first note that when $\hat \pi_\lambda^\dag(X)  \neq \pi_\lambda^\dag(X)$,  $\hat \tau(X) - \hat \beta \widehat{\textup{THR}}_U(X)$ and $\tau(X) - \beta^\dag \textup{THR}_U(X))$ have opposite signs, which implies that 
\begin{equation} \label{eq-s1} 
\begin{split}
| \tau(X) - \beta^\dag {\textup{THR}}_U(X) | 
  \leq {}& \big |  \{\hat \tau(X) - \hat \beta \widehat{\textup{THR}}_U(X)\} - \{ \tau(X) - \beta^\dag \textup{THR}_U(X))  \}     \big |  \\
={}& \big | \{\hat \tau(X) - \tau(X)\} - \{\hat \beta \widehat{\textup{THR}}_U(X) - \beta^\dag \textup{THR}_U(X) \}  \big |.    
  \end{split}
  \end{equation} 
  Then, we have that 
  \begin{align*}
  & U(\pi^\dag_\lambda) - U(\hat \pi_\lambda) \\
  ={}& \E\Big [   | \tau(X) - \beta^\dag {\textup{THR}}_U(X) | \cdot \I( \hat \pi_\lambda^\dag(X)  \neq \pi_\lambda^\dag(X)) \Big ]    \\
  \leq{}& \E\Big [  | \{\hat \tau(X) - \tau(X)\} - \{\hat \beta \widehat{\textup{THR}}_U(X) - \beta^\dag \textup{THR}_U(X) \} |  \cdot \I( \hat \pi_\lambda^\dag(X)  \neq \pi_\lambda^\dag(X)) \Big ] \\
  \leq{}&  || \{\hat \tau(X) - \tau(X)\} - \{\hat \beta \widehat{\textup{THR}}_U(X) - \beta^\dag \textup{THR}_U(X) \} ||_{\infty} \cdot  \E\Big [ \I( \hat \pi_\lambda^\dag(X)  \neq \pi_\lambda^\dag(X)) \Big ] \\
  \leq{}& || \{\hat \tau(X) - \tau(X)\} - \{\hat \beta \widehat{\textup{THR}}_U(X) - \beta^\dag \textup{THR}_U(X) \} ||_{\infty} \cdot \\
  {}& \quad \E\Big [ \I \Big ( | \tau(X) - \beta^\dag {\textup{THR}}_U(X) | \leq  | \{\hat \tau(X) - \tau(X)\} - \{\hat \beta \widehat{\textup{THR}}_U(X) - \beta^\dag \textup{THR}_U(X) \} | \Big )  \Big ] \\
  \leq{}& || \{\hat \tau(X) - \tau(X)\} - \{\hat \beta \widehat{\textup{THR}}_U(X) - \beta^\dag \textup{THR}_U(X) \} ||_{\infty}^{2}, 
  \end{align*} 
  where the first equality follows from Lemma \ref{lem-s3}, the first  inequality follows from equation \eqref{eq-s1}, the third inequality follows from the fact that the event $\{  \hat \pi_\lambda^\dag(X)  \neq \pi_\lambda^\dag(X) \}$ implies the event $\Big \{ \Big  | \tau(X) - \beta^\dag {\textup{THR}}_U(X) \Big | \leq \Big  | (\hat \tau(X) - \tau(X))  - (\hat \beta \widehat{\textup{THR}}_U(X) - \beta^\dag \textup{THR}_U(X) ) \Big | \Big \}$, and the last inequality follows from Lemma \ref{lem-s2}. 
  
  \hfill $\Box$

    \begin{lemma}[Local Strict Monotonicity] \label{lem-s5} Under Condition \ref{cond1},  there exists a constant $\epsilon_0 > 0$, such that for any $\beta \in \mathcal{B}(\beta^\dag, \epsilon_0) := \{\beta:  | \beta - \beta^\dag | \leq \epsilon \}$,   
  \[ \kappa \cdot | \beta - \beta^\dag | \leq   2 | \Psi(\beta; \bm{\eta}) - \Psi(\beta^\dag; \bm{\eta}) |,   \]
  where $\kappa$ is a constant defined in Condition \ref{cond1}. 
  \end{lemma}  
  
    \noindent 
  \emph{Proof of Lemma \ref{lem-s5}}.   
  By Condition \ref{cond1}(iii), 
  $|\partial \Psi(\beta^\dag; \bm{\eta})/\partial \beta | \geq \kappa > 0$ for a constant $\kappa$.  
  Since $\Psi(\beta; \bm{\eta}) $ is continuously differentiable with respect to $\beta$ around $\beta^\dag$ (Condition \ref{cond1}(i)). Thus, there exists a constant $\epsilon_0 > 0$, for all $\beta \in  \mathcal{B}(\beta^\dag, \epsilon_0)$, 
  \[       | \partial \Psi(\beta; \bm{\eta})/\partial \beta |  \geq \frac{\kappa}{2}.   \]
  Thus, by Taylor expansion, 
  \[   2  | \Psi(\beta; \bm{\eta}) - \Psi(\beta^\dag; \bm{\eta}) | = 2 | \partial \Psi(\bar \beta; \bm{\eta})/\partial \beta | \cdot | \beta - \beta^\dag  | \geq 2 \cdot \frac{\kappa}{2} \cdot | \beta - \beta^\dag |, \]
  where $\bar \beta$ is a point that lies between $\beta$ and $\beta^\dag$.        
  This finishes the proof. 
  
  \hfill $\Box$

    Lemma \ref{lem-s5} essentially indicates that the population estimating equation $\Psi(\beta; \eta) $ is a strictly monotonic function of $\beta$ around $\beta^\dag$.

  \medskip  
  \begin{lemma} \label{lem-s6}   Under Conditions \ref{cond1}-\ref{cond2}, there exists a constant $C_2 > 0$ such that 
  \[   | \hat \beta  - \beta^\dag | \leq  C_2  \cdot  || \hat \tau(X) - \tau(X) ||_{\infty}  \vee   || \widehat{\textup{THR}}_U(X) -  \textup{THR}_U(X) ||_{\infty} \vee N^{-1/2},    \]
  where $a \vee b = \max\{a,  b\}$.       
  \end{lemma}

  \noindent 
  \emph{Proof of Lemma \ref{lem-s6}}.  Recall that 
  \begin{align*}
  \Psi_n(\beta; \hat{\bm{\eta}}) ={}& n^{-1} \sum_{i=1}^n \widehat{\text{THR}}_U(X_i) \cdot  \I(\hat \tau(X_i) - \beta \cdot \widehat{\text{THR}}_U(X_i) > 0) - \lambda \\
  \Psi(\beta; \bm{\eta}) ={}&    \E[ \text{THR}_U(X) \cdot  \I(\tau(X) - \beta \cdot \text{THR}_U(X) > 0)] - \lambda,  
  \end{align*}
  $\Psi_n(\hat \beta; \hat{\bm{\eta}}) = 0$, and $\Psi(\beta^\dag; \bm{\eta}) = 0$.   We define $\tilde \beta$ as the solution of  
  \[     \Psi_n(\beta; \bm{\eta}) = 0.   \]

  We first prove that 
  \begin{align}  
  \tilde \beta  -  \beta^\dag ={}&  O_{\P}( n^{-1/2} ). \label{eq-s2}
  \end{align} 
  Observe that 
  \begin{align*}
  0 ={}& \Psi_n(\tilde \beta; \bm{\eta}) -   \Psi(\beta^\dag; \bm{\eta})  \\  
  ={}&   \{  \Psi_n(\tilde \beta; \bm{\eta}) - \Psi(\tilde \beta; \bm{\eta}) \}  + \{ \Psi(\tilde \beta; \bm{\eta})   - \Psi(\beta^\dag; \bm{\eta}) \}. 
  \end{align*}  
  Let 
  $\mathcal{F} :=  \{  \text{THR}_U(X) \cdot  \I(\tau(X) - \beta \cdot \text{THR}_U(X) > 0): \beta \in [0, M] \}$, 
  then $\sup_{\beta} |  \Psi_n( \beta;  \bm{\eta}) - \Psi(\beta;  \bm{\eta})  | = n^{-1/2}  \cdot || \mathbb{G}_n  ||_{\mathcal{F}}$, where $\mathbb{G}_n =\sqrt{n} (\P_n - \P)$. 
  Note that $\mathcal{F}$ is a class of monotone functions, then by Theorem \ref{thm2}.7.5 of \citet{vdv-1996}, 
  \[  \log N_{[]}(\epsilon, \mathcal{F}, L_2(\P)) \lesssim \frac{1}{\epsilon},    \]
  where $N_{[]}(\epsilon, \mathcal{F}, L_2(\P))$ is the bracketing number.      
  $\mathcal{F}$ has a finite bracketing integral.  In addition, $\mathcal{F}$ is also a class of bounded functions (bounded by 1). By Lemma 3.4.2 of  \citet{vdv-1996}, 
  \begin{align*}
  \E || \mathbb{G}_n  ||_{\mathcal{F}} \lesssim \tilde J_{[]}(1, \mathcal{F}, L_2(\P)) \left (1 + \frac{\tilde J_{[]}(1, \mathcal{F}, L_2(\P)) }{\sqrt{n}} \right ), 
  \end{align*}
  where $\tilde J_{[]}(1, \mathcal{F}, L_2(\P)) = \int_{0}^1 \sqrt{1+ \log N_{[]}(\epsilon, \mathcal{F}, L_2(\P))} d\epsilon$ be the bracketing integral of $\mathcal{F}$. Since 
  \begin{align*}
  \tilde J_{[]}(1, \mathcal{F}, L_2(\P)) 
  \lesssim{}& \int_{0}^1 \sqrt{1 +  \frac{1}{\epsilon}} d\epsilon  \\
  ={}&   \int_{1}^\infty \frac{\sqrt{1+t}}{t^2} dt \\
  \leq{}&  \int_{1}^\infty \frac{\sqrt{2t}}{t^2} dt < \infty. 
  \end{align*}
  then $\E || \mathbb{G}_n  ||_{\mathcal{F}} $ 
    is bounded, which implies that 
  $\sup_{\beta} |  \Psi_n( \beta;  \bm{\eta}) - \Psi(\beta;  \bm{\eta})  | =  O_{\P}( n^{-1/2})$ and  consequently, $\Psi_n(\tilde \beta; \bm{\eta}) - \Psi(\tilde \beta; \bm{\eta}) = O_{\P}( n^{-1/2})$. 
In addition, by Lemma \ref{lem-s5}, there exists a constant $\kappa$, such that  $\kappa | \tilde \beta - \beta^\dag | \leq 2 |\Psi(\tilde \beta; \bm{\eta}) - \Psi(\beta^\dag; \bm{\eta} ) |$. Thus, 
equation \eqref{eq-s2} holds.      

Then, we analyze $\hat \beta  -  \beta^\dag$. Consider the following decomposition,  
\begin{align*}
0   ={}& \Psi_n(\hat\beta; \hat{\bm{\eta}}) - \Psi_n(\tilde \beta; \bm{\eta}) \\
={}& \{  \Psi_n(\hat\beta; \hat{\bm{\eta}}) -  \Psi(\hat\beta; \hat{\bm{\eta}}) \} - \{ \Psi_n(\tilde \beta; \bm{\eta}) - \Psi(\tilde \beta; \bm{\eta})  \}   + \{ \Psi(\hat\beta; \hat{\bm{\eta}}) - \Psi(\tilde \beta; \bm{\eta})  \}.  
\end{align*}  
In the term $\Psi_n(\hat\beta, \hat{\bm{\eta}}) -  \Psi(\hat\beta, \hat{\bm{\eta}})$,  we can treat $\hat{\bm{\eta}}(X)$ as a fixed function of $X$ due to sample splitting. Then by a similar proof for \eqref{eq-s2}, we have  $\Psi_n(\tilde \beta; \bm{\eta}) - \Psi(\tilde \beta; \bm{\eta}) =  O_{\P}(n^{-1/2})$ and 
\[       \Psi_n(\hat\beta; \hat{\bm{\eta}}) -  \Psi(\hat\beta; \hat{\bm{\eta}}) = O_{\P}(n^{-1/2}).     \]
This yields that  
$\Psi(\hat\beta; \hat{\bm{\eta}}) - \Psi(\tilde \beta; \bm{\eta}) = O_{\P}(n^{-1/2}).$

  For $\Psi(\hat\beta; \hat{\bm{\eta}}) - \Psi(\tilde \beta; \bm{\eta})$, we further decompose it as follows: 
  \begin{equation} \label{eq-s3} 
\begin{split}
O_{\P}(n^{-1/2}) ={}& \Psi(\hat\beta; \hat{\bm{\eta}}) - \Psi(\tilde \beta; \bm{\eta}) \\
={}& \{  \Psi(\hat\beta; \hat{\bm{\eta}}) - \Psi(\hat \beta; \bm{\eta}) \} + \{ \Psi(\hat \beta; \bm{\eta})  -   \Psi(\tilde \beta; \bm{\eta}) \} \\ 
={}&   \{  \Psi(\hat\beta; \hat{\bm{\eta}}) - \Psi(\hat \beta; \bm{\eta}) \} + \{ \Psi(\hat \beta; \bm{\eta})  -  \Psi(\beta^\dag; \bm{\eta}) \}  + \{ \Psi(\beta^\dag; \bm{\eta}) -  \Psi(\tilde \beta; \bm{\eta}) \}      
\end{split} \end{equation}
For the first term on the right side of equation \eqref{eq-s3},      
\begin{align*} 
&\Psi(\hat\beta; \hat{\bm{\eta}}) - \Psi(\hat \beta; \bm{\eta}) \\
={}&        \E  \Big [  \widehat{\text{THR}}_U(X) \cdot \I(  \hat \tau(X) - \hat \beta \cdot \widehat{\text{THR}}_U(X) > 0 ) -   \textup{THR}_U(X)  \cdot \I(  \hat \tau(X) - \hat  \beta \cdot \widehat{\text{THR}}_U(X) > 0 ) \Big ],  \\
{}& +   \E  \Big [   \textup{THR}_U(X)  \cdot \I(  \hat \tau(X) - \hat \beta \cdot \widehat{\text{THR}}_U(X) > 0 ) -   \textup{THR}_U(X)  \cdot \I(   \tau(X) - \hat \beta \cdot \textup{THR}_U(x)  > 0 )           \Big ]  \\
\lesssim{}&  \sup_x | \widehat{\text{THR}}_U(X) - \textup{THR}_U(x)  |  +   \sup_{\beta}  \sup_x \big |  \hat \tau(x) - \tau(x) - \beta \cdot (\widehat{\text{THR}}_U(X) - \textup{THR}_U(x)  ) \big | \\
\lesssim{}&    || \hat \tau(X) - \tau(X) ||_{\infty}  \vee   || \widehat{\textup{THR}}_U(X) -  \textup{THR}_U(x) ||_{\infty}.
\end{align*}
For the third term $\Psi(\hat \beta; \bm{\eta}) -    \Psi(\tilde \beta; \bm{\eta})$ on the right side of equation \eqref{eq-s3}, by a Taylor expansion, 
\begin{align*} 
\Psi(\beta^\dag; \bm{\eta}) -    \Psi(\tilde \beta; \bm{\eta})
={}&     O_{\P} (  \tilde \beta - \beta^\dag ) = O_{\P} (n^{-1/2}).    
\end{align*}     
In addition, by Lemma \ref{lem-s5},  $\kappa | \hat \beta - \beta^\dag | \leq 2 |\Psi(\hat \beta; \bm{\eta}) - \Psi(\beta^\dag; \bm{\eta} ) |$. 
Therefore, we have 
\begin{align*}  
| \hat \beta  - \beta^\dag | \leq{}& C_2  \cdot   n^{-1/2} \vee  || \hat \tau(X) - \tau(X) ||_{\infty}  \vee   || \widehat{\textup{THR}}_U(X) -  \textup{THR}_U(X) ||_{\infty}   
\end{align*}
for a constant $C_2$. 
This implies the conclusion of Lemma \ref{lem-s6} by noting that $n = N/K$ and $K$ is a fixed integer. 

\hfill $\Box$

  \bigskip 
Finally, we prove Theorem \ref{thm3}. 

\noindent 
\emph{Proof of Theorem \ref{thm3}}.   
By Lemma \ref{lem-s4}, there exists a constant $C$, such that 
\begin{align*}    U(\pi_\lambda^\dag) - U(\hat \pi_\lambda)  \leq{}&  C \cdot ||  \{\hat \tau(X) - \tau(X)\} - \{\hat \beta \widehat{\textup{THR}}_U(X) - \beta^\dag \textup{THR}_U(X) \}  ||_{\infty}^{2} \\ 
\leq{}&  C_3  \cdot  || \hat \tau(X) - \tau(X) ||_{\infty}^2  \vee   || \widehat{\textup{THR}}_U(X) -  \textup{THR}_U(X) ||_{\infty}^2 \vee |\hat \beta - \beta^\dag|^2, 
\end{align*}
where $C_3$ is a constant.    
By Lemma \ref{lem-s6}, there exists a constant $C_2$, such that 
\[   | \hat \beta  - \beta^\dag | \leq  C_2  \vee  || \hat \tau(X) - \tau(X) ||_{\infty}  \vee   || \widehat{\textup{THR}}_U(X) -  \textup{THR}_U(X) ||_{\infty} \vee N^{-1/2}.     \]
Thus,  the conclusion holds under Condition \ref{cond3}. 

\hfill $\Box$

\subsection{Proof of Lemma \ref{lem2}}

\emph{Proof of Lemma \ref{lem2}}. By the definition of $\rho(x)$, under Assumption \ref{assump1}, we have 
\begin{align*}
\rho(x) ={}& \frac{ \E[ Y^0 Y^1 \mid X =x ] - \E[ Y^0 \mid X=x ] \cdot  \E[ Y^1 \mid X=x ] } {\sqrt{ \V(Y^0 \mid X=x) \cdot  \V(Y^1 \mid  X=x) } } \\
={}&    \frac{\P(Y^0=1, Y^1=1 \mid X=x) - \mu_0(x)\mu_1(x) }{\sqrt{  \mu_0(x)(1-\mu_0(x) ) \mu_1(x)(1-\mu_1(x))  }}.
\end{align*} 
Applying Frechet--Hoeffding inequality to $\P(Y^0=1, Y^1=1 \mid X=x)$ yields that 
\begin{align*}
\P(Y^0=1, Y^1=1 \mid X=x) \geq{}& \max\{\mu_0(x) + \mu_1(x) -1, 0 \} \\
\P(Y^0=1, Y^1=1 \mid X=x) \leq{}& \min\{\mu_0(x),  \mu_1(x)\},
\end{align*}  
which implies 
       		\begin{align*}
	\rho(x) \geq  L_{\rho}(x)  
        ={}&   \frac{ \max\{ \mu_0(x) + \mu_1(x) - 1 - \mu_0(x) \mu_1(x), -\mu_0(x)\mu_1(x)\} }{\sqrt{  \mu_0(x)(1-\mu_0(x) ) \mu_1(x)(1-\mu_1(x))  }}       \\
	  ={}& - \frac{ \min\{(1-\mu_0(x))(1-\mu_1(x)), \mu_0(x)\mu_1(x)\} }{\sqrt{  \mu_0(x)(1-\mu_0(x) ) \mu_1(x)(1-\mu_1(x))  }}	\end{align*}
and 
\[  \rho(x) \leq 
	 U_{\rho}(x) =  \frac{ \min\{\mu_0(x)(1-\mu_1(x)), \mu_1(x)(1-\mu_0(x)) \} }{\sqrt{  \mu_0(x)(1-\mu_0(x) ) \mu_1(x)(1-\mu_1(x))  }}.  
 \]	   
 This finishes the proof. 
  
\hfill $\Box$

\subsection{Proof of Proposition \ref{prop2}}

\emph{Proof of Proposition \ref{prop2}}.  
First, we prove that $$ \max \Big\{  \mu_0(x) \{ 1 - \mu_1(x) \} -    \bar \rho(x)  \sqrt{  \mu_0(x)\{1-\mu_0(x) \} \mu_1(x)\{1-\mu_1(x) \}    },  0  \Big\}$$ is an upper bound on $\text{THR}(x)$.  This follows immediately from the fact that  $\text{THR}(x) \geq 0$ by its definition, and 
    \[  \rho(x) =   \frac{\mu_0(x)\{1-\mu_1(x)\} - \text{THR}(x) }{  \sqrt{  \mu_0(x)\{1-\mu_0(x) \} \mu_1(x)\{1-\mu_1(x) \}    } }     \geq \bar\rho(x) :=  \max\{ \tilde \rho_L(x), \rho_L(x)\}     \]
by Assumption \ref{assump3} and Lemma \ref{lem2}.

Then, we show that this upper bound is sharp. When $\bar \rho(x) = \rho_L(x)$ (i.e., $\tilde \rho_L(x) \leq \rho_L(x)$),  the upper bound reduces to the Frechet--Hoeffding upper bound under Assumption \ref{assump1}, that is, 
      \begin{align*}
 & \max \Big\{  \mu_0(x) \{ 1 - \mu_1(x) \} -    \bar \rho(x)  \sqrt{  \mu_0(x)\{1-\mu_0(x) \} \mu_1(x)\{1-\mu_1(x) \}    },  0  \Big\}\\
 = {}&     \min\{ \mu_0(x),  1- \mu_1(x) \}  \end{align*}
 The Frechet--Hoeffding upper bound is achievable if $\P(Y^0=1, Y^1 = 1| X=x) = 0$ or $\P(Y^0=0, Y^1 = 0 | X=x) = 0$. This is because 
    \begin{align*}
  & \P(Y^0=1, Y^1 = 1| X=x) = 0 \\ 
  \iff{}& \mu_0(x) = \P(Y^0 = 1 | X =x) = \P(Y^0 = 1, Y^1 = 0 | X =x) = \text{THR}(x), 
\end{align*} 
and 
 \begin{align*}
  & \P(Y^0=0, Y^1 = 0 | X=x) = 0 \\
  \iff{}&   1 - \mu_1(x) = \P(Y^1 = 0 | X =x) = \P(Y^0 = 1, Y^1 = 0 | X =x) = \text{THR}(x).
 \end{align*}

When $\bar \rho(x) = \tilde \rho_L(x)$  (i.e., $\tilde \rho_L(x) > \rho_L(x)$),  the upper bound is achievable  if $\rho(x) =  \tilde \rho_L(x)$. 
This finishes the proof.

\hfill $\Box$

 \section{Conservative Strategy}  \label{sec4-1-3}In this section, we also propose a new alternative estimation strategy, which is provided in the main text as a comparative method for the simulation studies. 
 The pessimistic strategy of controlling the worst-case may be overly conservative,
 as it uniformly applies the upper bound $\text{THR}_U(x)$ across all $x$.
  In fact, the upper bound $\text{THR}_U(x)$ is attained if and only if $\P(Y(0)=1, Y(1) = 1\mid X=x) = 0$ or $\P(Y(0)=0, Y(1) = 0 \mid X=x) = 0$. These conditions are satisfied only under extreme data-generating processes. Therefore, it is practically necessary to develop less pessimistic strategies. 

 Instead of controlling the worst-case, a slightly compromised strategy is to control the worst-case with a high probability. To achieve this, we define $\text{THR}_{\alpha}$ as the $(1-\alpha)$-th quantile of $\text{THR}_U(X)$, i.e., $\P( \text{THR}_U(X) \leq  \text{THR}_{\alpha} ) \geq 1 - \alpha.$  
We then replace $\text{THR}_U(x)$ with $\text{THR}_{\alpha}(x) := \min\{  \text{THR}_{\alpha},  \text{THR}_U(x) \}$ 
and use the same estimation procedures as the pessimistic strategy to estimate the corresponding ITRs. 
This strategy is equivalent to estimating the optimal ITR that solves the constrained optimization: 
 $$\max_{\pi}~  R(\pi), s.t.~  \text{THR}_{\alpha}(\pi) \leq \lambda,$$ 
where $\text{THR}_{\alpha}(\pi) = \E[ \text{THR}_{\alpha}(X)  \pi(X)]$. 
Since the upper bound $\text{THR}_U(X)$ is less than or equal to $\text{THR}_{\alpha}(X)$ with probability at least $1 - \alpha$, the constraint $\text{THR}{\alpha}(\pi) \leq \lambda$ ensures that the true harm rate $\text{THR}(\pi)$ is below $\lambda$ with probability at least $1 - \alpha$. For practical applications, we may set $\alpha$ to 0.05 or 0.10.

\section{Additional Results for Simulation}

Figure \ref{fig-s1} displays the Boxplot of the harm rate and the reward induced by the estimated ITRs for different methods, using a harm rate threshold of $\lambda = 0.02$.  
Figure \ref{fig-s1} shows similar patterns to those in Figure 1 in the manuscript.  

\begin{figure}[!h]
\centering
\subfloat[$\lambda = 0.02$, {\bf without cross fitting}]{
\begin{minipage}[t]{0.5 \linewidth} 
\centering
\includegraphics[width=1\textwidth]{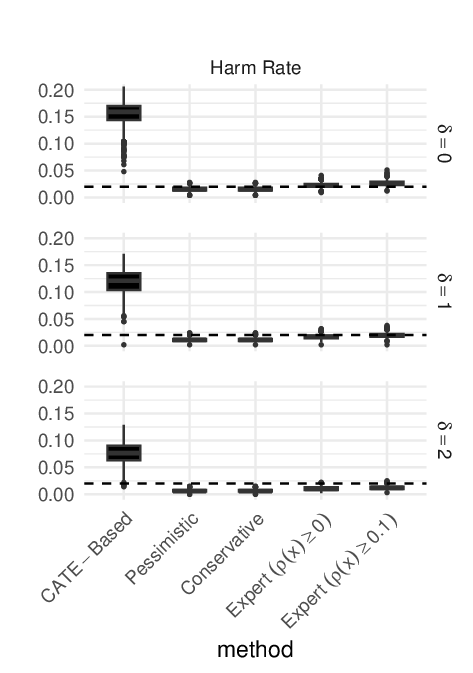}
\end{minipage}%
} \hspace{-20pt} 
\subfloat[$\lambda = 0.02$, {\bf without cross fitting}]{
\begin{minipage}[t]{0.5 \linewidth}
\centering
\includegraphics[width=1\textwidth]{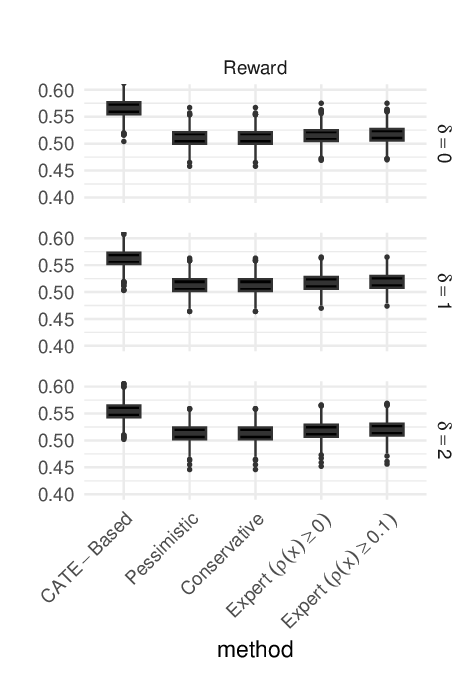}
\end{minipage} 
} 
\caption{(a) Boxplot of the treatment harm rate induced by estimated ITRs for various methods, where the dotted line indicates the threshold of harm rate ($\lambda = 0.02$); (b) Boxplot of the reward induced by the estimated ITR for various methods.} 
\label{fig-s1} 
\end{figure}

The results in Figure \ref{fig-s1} are obtained without cross-fitting. To assess robustness, we also present results using different cross-fitting folds. Figures \ref{fig-s2} and \ref{fig-s3} show the results for 5-fold and 10-fold cross-fitting, respectively. These results are  similar to those  those in Figure \ref{fig-s1}, indicating that the proposed methods perform stably across different folds.

\begin{figure}[H]
\centering
\subfloat[ $\lambda = 0.02$, {\bf with five-fold cross fitting}]{
\begin{minipage}[t]{0.5 \linewidth} 
\centering
\includegraphics[width=1\textwidth]{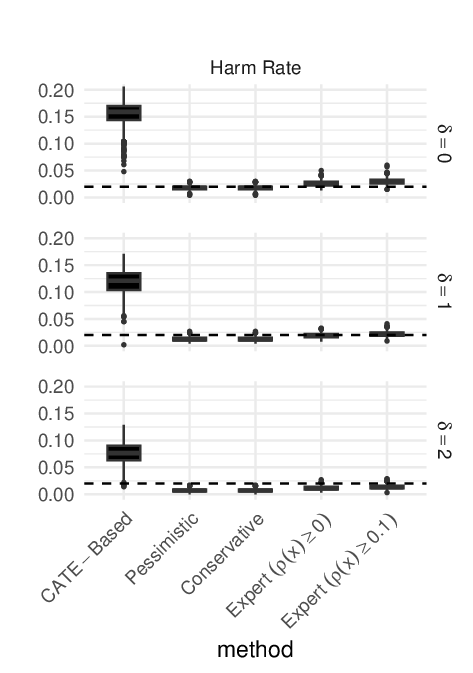}
\end{minipage}%
} \hspace{-20pt} 
\subfloat[$\lambda = 0.02$, {\bf with five-fold cross fitting}]{
\begin{minipage}[t]{0.5 \linewidth}
\centering
\includegraphics[width=1\textwidth]{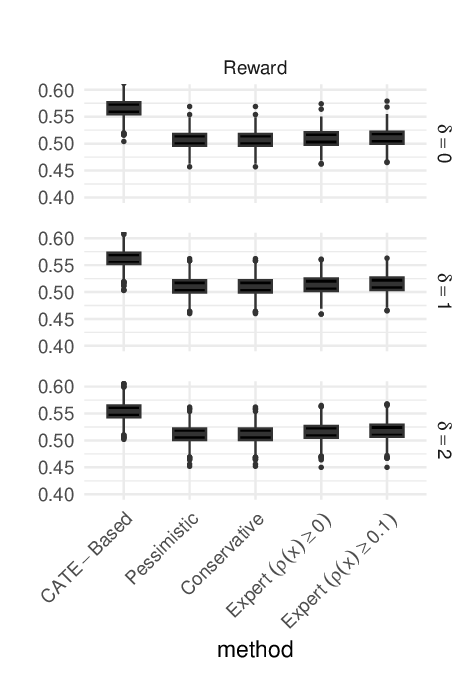}
\end{minipage} 
} 
\caption{(a) Boxplot of the treatment harm rate induced by estimated ITRs for various methods, where the dotted line indicates the threshold of harm rate ($\lambda = 0.02$); (b) Boxplot of the reward induced by the estimated ITR for various methods.} 
\label{fig-s2} 
\end{figure}

\begin{figure}[H]
\centering
\subfloat[ $\lambda = 0.02$, {\bf with ten-fold cross fitting}]{
\begin{minipage}[t]{0.5 \linewidth} 
\centering
\includegraphics[width=1\textwidth]{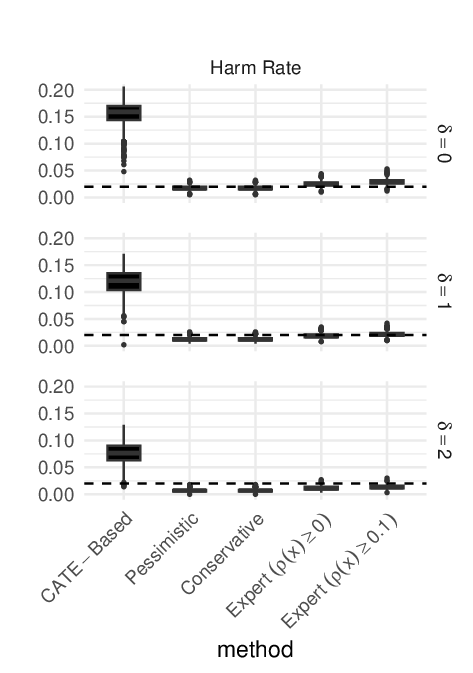}
\end{minipage}%
} \hspace{-20pt} 
\subfloat[$\lambda = 0.02$, {\bf with ten-fold cross fitting}]{
\begin{minipage}[t]{0.5 \linewidth}
\centering
\includegraphics[width=1\textwidth]{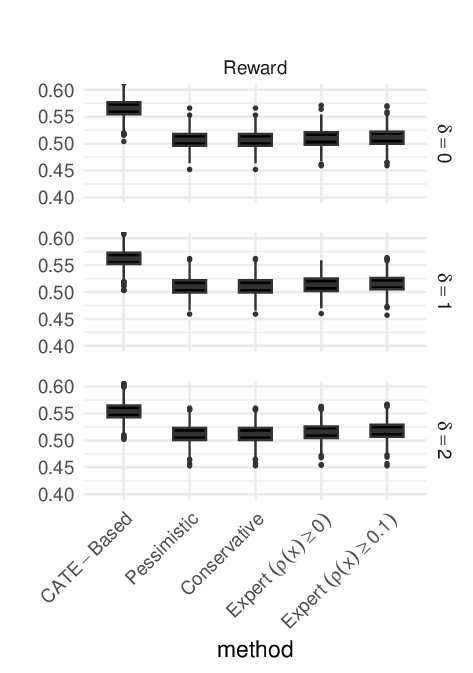}
\end{minipage} 
} 
\caption{(a) Boxplot of the treatment harm rate induced by estimated ITRs for various methods, where the dotted line indicates the threshold of harm rate ($\lambda = 0.02$); (b) Boxplot of the reward induced by the estimated ITR for various methods.} 
\label{fig-s3} 
\end{figure}

\newpage 
\section{Further Discussion on Optimization Problem}
In extreme risk-averse application scenarios, 
it may be of interest to control the harm rate as much as possible while ensuring that the benefit rate exceeds certain thresholds. This problem can be formulated as estimating the ITR that satisfies 
\begin{align}  \label{eq-s5}
    \begin{cases}
             \min_{\pi} \quad \text{THR}(\pi)  \\
s.t. \quad  R(\pi) \geq \lambda.         
    \end{cases}
\end{align} 


Similar to Theorem 1 in the manuscript, the optimal ITR for the above optimization problem has an explicit form.

\begin{theorem}[Optimal ITR] \label{thm-s1}

For the optimization problem \eqref{eq-s5}, 

(a) if $\lambda \leq \E[ Y(0) ]$, an optimal ITR is $\pi_\lambda^*(X) \equiv 0$. In this case, $\textup{THR}(\pi_\lambda^*) = 0$.  


(b) if $\E[ \tau(X) \cdot \mathbb{I}(\textup{THR}(X) = 0)] > 0$, 
  
  \begin{itemize}
  	\item  if $\E[ Y(0) ] < \lambda  \leq \E[Y(0)] + \E[ \tau(X) \cdot \mathbb{I}(\textup{THR}(X) = 0) ]$,  an  optimal ITR is $\pi_\lambda^*(X) = \mathbb{I}( \textup{THR}(X) = 0 )$.  In this case, $\textup{THR}(\pi_\lambda^*) = 0$.  
	
	\item  if $\E[Y(0)] + \E[ \tau(X) \cdot \mathbb{I}(\textup{THR}(X) = 0) ] <  \lambda \leq \E[Y(0)]  +  \E[ \tau(X) \cdot \mathbb{I}(\tau(X) > 0) ]$, the optimal ITR is $\pi_\lambda^*(X) = \mathbb{I}( \beta^* \tau(X) -   \textup{THR}(X) > 0 )$, where $\beta^*$ satisfies 
\[ \E[Y(0)] +  \E[ \tau(X) \cdot \I( \beta^* \tau(X) -   \textup{THR}(X)  > 0)] = \lambda. \]

 	\item  if $ \E[Y(0)]  +  \E[ \tau(X) \cdot \mathbb{I}(\tau(X) > 0) ] < \lambda$, no solution exists.
  
  \end{itemize}

(c) if $\E[ \tau(X) \cdot \mathbb{I}(\textup{THR}(X) = 0)] \leq 0$, 

\begin{itemize}
	\item  if $\E[Y(0)]  <  \lambda \leq \E[Y(0)]  +  \E[ \tau(X) \cdot \mathbb{I}(\tau(X) > 0) ]$, the optimal ITR is $\pi_\lambda^*(X) = \mathbb{I}( \beta^* \tau(X) -   \textup{THR}(X) > 0 )$, where $\beta^*$ satisfies 
\[ \E[Y(0)] +  \E[ \tau(X) \cdot \I( \beta^* \tau(X) -   \textup{THR}(X)  > 0)] = \lambda. \]

 	\item  if $ \E[Y(0)]  +  \E[ \tau(X) \cdot \mathbb{I}(\tau(X) > 0) ] < \lambda$, no solution exists.
\end{itemize}

\end{theorem} 

\noindent 
\emph{Proof of Theorem \ref{thm-s1}}.  
 Theorem \ref{thm-s1}(a) is straightforward. We now proceed to prove Theorem \ref{thm-s1}(b); the proof of Theorem \ref{thm-s1}(c) follows similarly. 
   \begin{table}[h!] 
\centering
  \centering
  \begin{tabular}{cccc}
    \toprule
                 &  $\tau(X) > 0$ &  $\tau(X) \leq 0 $ \\
    \midrule 
     $\text{THR}(X) > 0$  & $\mathcal{G}_1$ & $\mathcal{G}_2$ \\
    $\text{THR}(X) = 0 $  & $\mathcal{G}_3$ & $\mathcal{G}_4$ \\
    \bottomrule
  \end{tabular}
\end{table}

  When $\E[ \tau(X) \cdot \mathbb{I}(\textup{THR}(X) = 0)] > 0$, 

\begin{itemize}
    \item if $\E[ Y(0) ] < \lambda  \leq \E[Y(0)] + \E[ \tau(X) \cdot \mathbb{I}(\textup{THR}(X) = 0) ]$,
     then we can simply assign treatment to all units for which $\textup{THR}(X) = 0$. This ITR can be represented as $\pi_\lambda^*(X) = \mathbb{I}( \textup{THR}(X) = 0 )$.  It yields a  reward greater than $\lambda$, while satisfying $\textup{THR}(\pi_\lambda^*) = 0$

    \item  if $\E[Y(0)] + \E[ \tau(X) \cdot \mathbb{I}(\textup{THR}(X) = 0) ] <  \lambda \leq \E[Y(0)]  +  \E[ \tau(X) \cdot \mathbb{I}(\tau(X) > 0) ]$,
    then the optimal ITR involves 
a trade-off between reward and harm. By a similar proof of Theorem 1, we can show that  
    the optimal ITR is $\pi_\lambda^*(X) = \mathbb{I}( \beta^* \tau(X) -   \textup{THR}(X) > 0 )$, where $\beta^*$ satisfies 
\[ \E[Y(0)] +  \E[ \tau(X) \cdot \I( \beta^* \tau(X) -   \textup{THR}(X)  > 0)] = \lambda. \]

\item if $ \E[Y(0)]  +  \E[ \tau(X) \cdot \mathbb{I}(\tau(X) > 0) ] < \lambda$, then no ITR that satisfies the constraint in the optimization problem, and thus no solution exists. 
\end{itemize}


 \hfill $\Box$

\end{document}